\documentclass[apj,numberedappendix,revtex4]{emulateapj}
\usepackage[backref,breaklinks,colorlinks,citecolor=blue]{hyperref}
\usepackage{amssymb,amsmath}
\usepackage{graphicx}
\usepackage{appendix}
\usepackage{natbib}
\usepackage{enumerate}
\usepackage{multirow}
\usepackage{bm}

\newcommand{\lenstool}{{\tt{Lenstool}}}

\altaffiltext{\MIT}{Kavli Institute for Astrophysics and Space Research, Massachusetts Institute of Technology, 77 Massachusetts Avenue, Cambridge, MA 02139}
\altaffiltext{\Waterloo}{Department of Physics and Astronomy, University of Waterloo, Waterloo, ON, Canada}
\altaffiltext{\MSU}{Department of Physics and Astronomy, Michigan State University, East Lansing, MI 48824, USA}
\altaffiltext{\FNAL}{Fermi National Accelerator Laboratory, Batavia, IL 60510-0500, USA}
\altaffiltext{\KICPChicago}{Kavli Institute for Cosmological Physics, University of Chicago, Chicago, IL, USA 60637}
\altaffiltext{\AAUChicago}{Department of Astronomy and Astrophysics, University of Chicago, Chicago, IL, USA 60637}
\altaffiltext{\Miss}{Department of Physics and Astronomy, University of Missouri, 5110 Rockhill Road, Kansas City, MO 64110}
\altaffiltext{\KIPAC}{Kavli Institute for Particle Astrophysics and Cosmology, Stanford University, 452 Lomita Mall, Stanford, CA 94305}
\altaffiltext{\Stanford}{Department of Physics, Stanford University, 382 Via Pueblo Mall, Stanford, CA 94305}
\altaffiltext{\Goddard}{Observational Cosmology Lab, Goddard Space Flight Center, 8800 Greenbelt Road, Greenbelt, MD 20771, USA}
\altaffiltext{\Huntingdon}{Huntingdon Institute for X-ray Astronomy, LLC}
\altaffiltext{\Princeton}{Department of Astrophysical Sciences, Princeton University, 4 Ivy Lane, Princeton, NJ 08544-1001, USA}
\altaffiltext{\UMontreal}{D\'{e}partement de Physique, Universit\'{e} de Montr\'{e}al, C.P. 6128, Succ. Centre-Ville, Montr\'{e}al, Qu\'{e}bec H3C 3J7, Canada}
\altaffiltext{\UMD}{Department of Astronomy, University of Maryland, College Park, MD 20742, USA}
\altaffiltext{\Melbourne}{School of Physics, University of Melbourne, Parkville, VIC 3010, Australia}
\altaffiltext{\IoA}{Institute of Astronomy, Madingley Road, Cambridge CB3 0HA}
\altaffiltext{\Trieste}{Astronomy Unit, Department of Physics, University of Trieste, via Tiepolo 11, I-34131 Trieste, Italy}
\altaffiltext{\IFPU}{Institute for Fundamental Physics of the Universe, Via Beirut 2, 34014 Trieste, Italy}
\altaffiltext{\INAF}{INAF-Osservatorio Astronomico di Trieste, via G. B. Tiepolo 11, I-34143 Trieste, Italy}
\altaffiltext{\UMich}{Department of Astronomy, University of Michigan, 1085 S. University, Ann Arbor, MI 48109}
\altaffiltext{\CfA}{Harvard-Smithsonian Center for Astrophysics, 60 Garden Street, Cambridge, MA 02138, USA}
\altaffiltext{\Leiden}{Leiden Observatory, Leiden University, PO Box 9513, 2300 RA Leiden, The Netherlands}
\altaffiltext{$\dagger$}{Lyman Spitzer Jr. Fellow}

\def\MIT{1}
\def\Waterloo{2}
\def\MSU{3}
\def\FNAL{4}
\def\KICPChicago{5}
\def\AAUChicago{6}
\def\Miss{7}
\def\KIPAC{8}
\def\Stanford{9}
\def\Goddard{10}
\def\Huntingdon{11}
\def\Princeton{12}
\def\UMontreal{13}
\def\UMD{14}
\def\Melbourne{15}
\def\IoA{16}
\def\Trieste{17}
\def\IFPU{18}
\def\INAF{19}
\def\UMich{20}
\def\CfA{21}
\def\Leiden{22}

\begin{document}


\title{Anatomy of a Cooling Flow: \\The Feedback Response to Pure Cooling in the Core of the Phoenix Cluster}
   
\author{
M.~McDonald\altaffilmark{\MIT},
B.~R.~McNamara\altaffilmark{\Waterloo},
G.~M.~Voit\altaffilmark{\MSU},
M.~Bayliss\altaffilmark{\MIT},
B.\,A.\,Benson\altaffilmark{\FNAL,\KICPChicago,\AAUChicago},
M.\,Brodwin\altaffilmark{\Miss},
R.\ E.\ A.\ Canning\altaffilmark{\KIPAC,\Stanford},
M.~K.~Florian\altaffilmark{\Goddard},
G.~P.~Garmire\altaffilmark{\Huntingdon},
M.~Gaspari\altaffilmark{\Princeton,$\dagger$},
M.~D.~Gladders\altaffilmark{\KICPChicago,\AAUChicago},
J.~Hlavacek-Larrondo\altaffilmark{\UMontreal},
E.~Kara\altaffilmark{\MIT,\UMD},
C.~L.~Reichardt\altaffilmark{\Melbourne},
H.~R.~Russell\altaffilmark{\IoA},
A.~Saro\altaffilmark{\Trieste,\IFPU,\INAF},
K.~Sharon\altaffilmark{\UMich},
T.~Somboonpanyakul\altaffilmark{\MIT},
G.~R.~Tremblay\altaffilmark{\CfA},
R.~J.~van Weeren\altaffilmark{\Leiden}
}

\email{Email: mcdonald@space.mit.edu}   


\begin{abstract}

We present new, deep observations of the Phoenix cluster from the \emph{Chandra X-ray Observatory}, the \emph{Hubble Space Telescope}, and the Karl Jansky Very Large Array.
These data provide an order of magnitude improvement in depth and/or angular resolution at X-ray, optical, and radio wavelengths, yielding an unprecedented view of the core of the Phoenix cluster.
We find that the one-dimensional temperature and entropy profiles are consistent with expectations for pure-cooling hydrodynamic simulations and analytic descriptions of homogeneous, steady-state cooling flow models. In particular, the entropy profile is well-fit by a single power law at all radii, with no evidence for excess entropy in the core.
In the inner $\sim$10\,kpc, the cooling time is shorter by an order of magnitude than any other known cluster, while the ratio of the cooling time to freefall time ($t_{cool}/t_{ff}$) approaches unity, signaling that the ICM is unable to resist multiphase condensation on kpc scales.
When we consider the thermodynamic profiles in two dimensions, we find that the cooling is highly asymmetric. The bulk of the cooling in the inner $\sim$20\,kpc is confined to a low-entropy filament extending northward from the central galaxy, with $t_{cool}/t_{ff} \sim 1$ over the length of the filament. This northern filament is significantly absorbed, suggesting the presence of $\sim$10$^{10}$ M$_{\odot}$ in cool gas that is absorbing soft X-rays.
We detect a substantial reservoir of cool ($\sim$10$^4$\,K) gas (as traced by the [O\,\textsc{ii}]$\lambda\lambda$3726,3729 doublet), which is coincident with the low-entropy filament. 
The bulk of this cool gas is draped around and behind a pair of X-ray cavities, presumably bubbles that have been inflated by radio jets, which are detected for the first time on kpc scales. These data support a picture in which AGN feedback is promoting the formation of a multiphase medium via a combination of ordered buoyant uplift and locally enhanced turbulence. These processes ought to counteract the tendency for buoyancy to suppress condensation, leading to rapid cooling along the jet axis. The recent mechanical outburst has sufficient energy to offset cooling, and appears to be coupling to the ICM via a cocoon shock, raising the entropy in the direction orthogonal to the radio jets.



\end{abstract}

 \vspace{-0.2in}

\section{Introduction}
\setcounter{footnote}{0}

In the cores of some galaxy clusters, the intracluster medium (ICM) can be dense enough and cool enough that the cooling time reaches $\lesssim$1\,Gyr. 
The total mass of hot ($\gtrsim$10$^8$\,K) gas in these so-called ``cool cores'', divided by the short cooling time, implies cooling rates of $\sim$100--1000 M$_{\odot}$ yr$^{-1}$, which is considerably more than the typically-observed star formation rates (SFRs) of $\sim$1--10 M$_{\odot}$ yr$^{-1}$ in central brightest cluster galaxies \citep[BCGs;][]{mcnamara89,crawford99,edwards07,hatch07,odea08,mcdonald10,hoffer12,rawle12,donahue15,molendi16,mcdonald18a}. It is thought that mechanical feedback from the radio-loud central active galactic nucleus (AGN) is suppressing the bulk of the cooling \citep[e.g.,][]{mcnamara07,gaspari11,mcnamara12,fabian12,gaspari13,li14,prasad15,li17,yang19}, leading to SFRs that represent only $\sim$1\% of the predicted cooling rate \citep[e.g.,][]{odea08,mcdonald18a}.
Without this feedback, the ICM would cool rapidly \citep[i.e., ``cooling flows'';][]{fabian84}, leading to central galaxies that are much more massive than those observed today \citep[e.g.,][]{silk12,main17}. While the masses of present-day giant elliptical galaxies rule out 
cooling flows at sustained rates of hundreds of solar masses per year for periods of several Gyr, the details of the cooling/feedback cycle remain unclear, including whether short-duration runaway cooling can occur in clusters \citep[see recent work by][]{voit15,mcnamara16,gaspari18}.

While it has been known for some time that star formation is suppressed in central cluster galaxies, this suppression factor, its cluster-to-cluster scatter, and its dependence on cluster mass has only recently been constrained to high accuracy. Our recent work \citep{mcdonald18a} examined the correlation between star formation rate and classical cooling rate in a sample of 107 galaxies, galaxy groups, and galaxy clusters. With this large sample spanning a wide range in group/cluster masses we found that star formation was less suppressed in the most massive clusters and that 
the four clusters harboring the most rapidly-accreting central supermassive black holes (\.M$_{acc}$/\.M$_{Edd}$ $>$ 5\%) all have elevated central SFRs. We proposed that AGN feedback in these systems may be saturating -- the effectiveness of AGN feedback will naturally be limited by the Eddington luminosity. Further, at high accretion rates much (but not all) of the AGN power output is radiative, which may lead to \emph{cooling} of the ICM via the inverse Compton effect close to the AGN \citep[e.g.,][]{russell13}. As the cluster mass increases, the amount of gas available for cooling also increases. 
For the most massive clusters, even a cooling flow that is suppressed by a factor $\sim$100 can be close to the Eddington rate of the central black hole.
 
 The Phoenix cluster, first identified using the South Pole Telescope \citep[SPT-CLJ2344-4243;][]{carlstrom11,williamson11}, represents 
 a unique opportunity to study the potential limitations of AGN feedback. This system is the most X-ray luminous cluster known and harbors the most star-forming central galaxy known, with measurements of the star formation rate ranging from 500--800 M$_{\odot}$ yr$^{-1}$ \citep{mcdonald12c,mcdonald13a,mittal17}. The high X-ray luminosity and BCG star formation rate are most likely related, as the starburst appears to be fueled by the rapid cooling (high luminosity) of the ICM \citep{mcdonald12c}, implying that $>$10\% of the predicted cooling is ultimately realized in star formation.
Attempts to measure the cooling rate at intermediate temperatures, based on \emph{XMM-Newton} RGS spectroscopy, have suggested that a significant amount ($\sim$500 M$_{\odot}$ yr$^{-1}$) of the hot gas 
cools below the ambient central temperature of $\sim$2\,keV, and that this cooling is concentrated in the very inner region \citep{tozzi15b,pinto18}. At significantly lower temperatures, \cite{mcdonald14a} and \cite{russell17} have found extended, filamentary gas in the warm ionized and cold molecular phases, respectively. 
The Phoenix cluster is also unique in that it is one of only 4 known clusters harboring a central quasar \citep{russell13,ueda13}, while also having tremendously powerful radio jets \citep{hlavacek15,mcdonald15b}. This active galactic nucleus (AGN) appears to be 
supplying a power of $\sim$10$^{46}$ erg s$^{-1}$.  Roughly half of the power is radiative
(quasar-mode) feedback, while the other half is mechanical (radio-mode) feedback. This makes the central AGN in Phoenix one of the most powerful in the known Universe \citep[see review by][]{fabian12}.

In this work, we present new, deep data on the core of the Phoenix cluster from the \emph{Chandra X-ray Observatory}, \emph{Hubble Space Telescope}, and the Karl G.\ Jansky Very Large Array. These data provide an order of magnitude improvement in depth and/or angular resolution at X-ray, optical, and radio wavelengths, yielding a detailed view of the complex physics at the center of this cluster. In particular, the deep X-ray data allow us to carefully model the contribution to the X-ray emission from the bright point source, allowing us to map out the thermodynamics of the ICM on scales of $\lesssim$10\,kpc, where the bulk of the cool gas is observed. We present these new data, as well as the data reduction and analysis in \S2. In \S3 we summarize the first results from these new data, focusing on the reservoir of cool gas in the inner 30\,kpc (\S3.1), the one-dimensional thermodynamic profiles (\S3.2), and the two-dimensional thermodynamic maps (\S3.3). In \S4 we present a discussion of these findings, trying to place the Phoenix cluster in the context of the cooling/feedback loop observed in other nearby clusters. Finally, we conclude in \S5 with a summary of the key results and a brief preview of future work.

Throughout this work, we assume H$_0 = 70$ km s$^{-1}$ Mpc$^{-1}$, $\Omega_M = 0.27$, and $\Omega_{\Lambda} = 0.73$. We assume $z = 0.597$ for the Phoenix cluster, which is based on optical spectroscopy of the member galaxies and the central brightest cluster galaxy \citep{ruel14,mcdonald14a,bleem15}.

\section{Data}

\subsection{Optical: Hubble Space Telescope}

New optical data for this system were acquired from the \emph{Hubble Space Telescope} (HST) during program GO15315 (PI: McDonald). These data included broadband F475W (2 orbits), F775W (2 orbits), and F850LP (1 orbit), using the ACS/WFC camera.  In addition, we obtained 16 exposures over 8 orbits in the FR601N ramp filter, centered on 5952\AA, which corresponds to the position of the [O\,\textsc{ii}]$\lambda\lambda$3726,2729 doublet at $z=0.597$. 
All broadband and narrowband exposures were taken in a 4-point box dither pattern, with spacing of 0.262$^{\prime\prime}$ between points.

\begin{figure}[t!]
\centering
\includegraphics[width=0.48\textwidth]{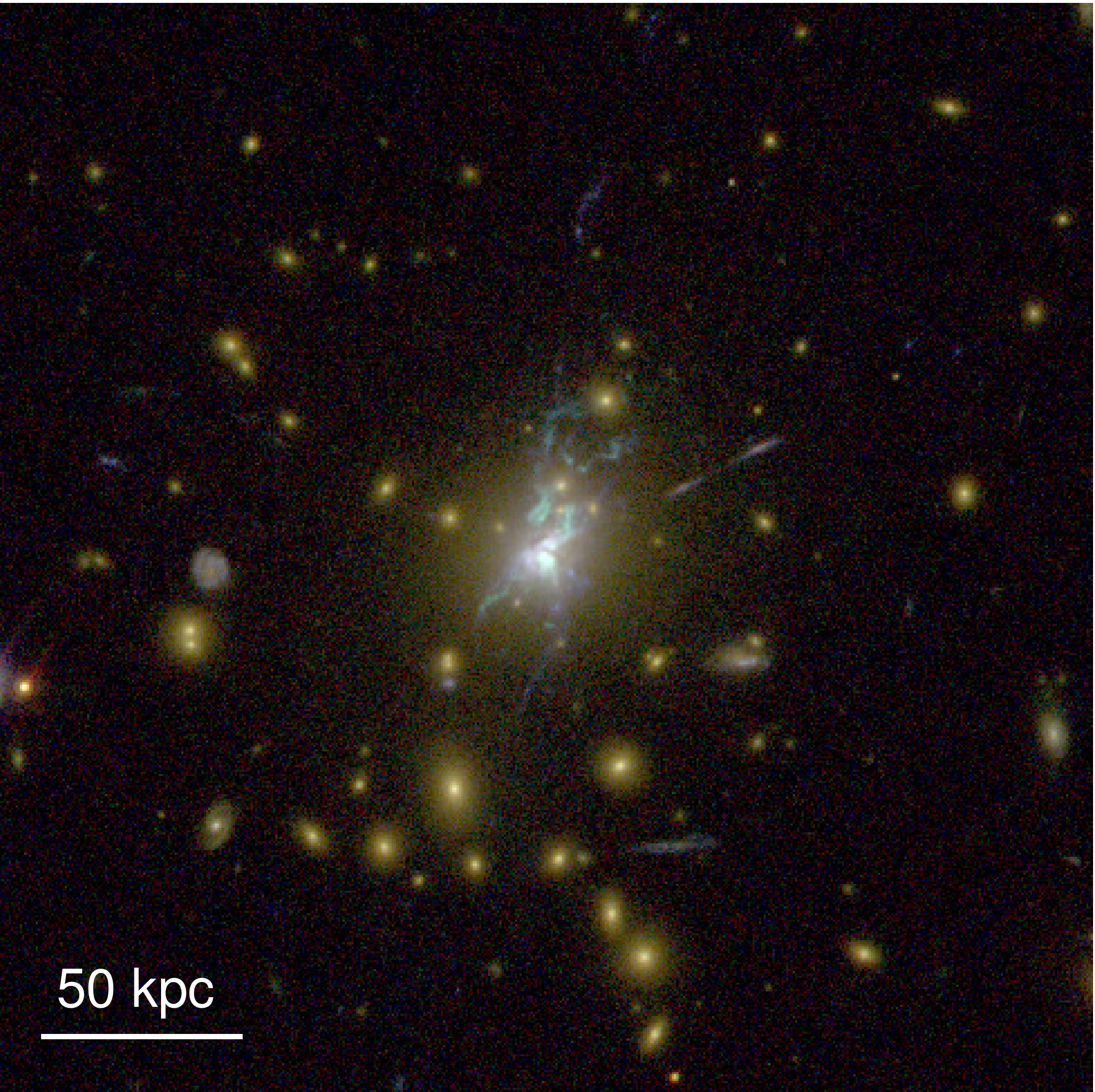}
\caption{\emph{Hubble Space Telescope} ACS-WFC2 color image of the core of the Phoenix cluster, made by combining images in the F475W, F775W, and F850LP bands. This image shows a giant elliptical central galaxy with a morphologically complex, dusty starburst component that extends for $\sim$50\,kpc to the north and south. Previously identified \citep{mcdonald15b} linear filaments to the north and northwest, which extend for $\sim$100\,kpc have been spectroscopically identified as radial arcs (see \S2.1).}
\label{fig:rgb}
\end{figure}

For each band, we used \textsc{AstroDrizzle} v2.2.4 with the default parameters to make a first, rough image free from cosmic rays. We then used \textsc{SExtractor} \citep{bertin96} to generate a list of point sources, which are then used with \textsc{tweakreg} to adjust the astrometry for each individual HST frame before recombining a second time with \textsc{AstroDrizzle}. For the ramp filter data, which has a very narrow field and low throughput, stars were outnumbered by cosmic rays by a factor of $>$1000:1, meaning that any attempt to automate this process would fail. For these frames, we identified by eye a set of 27 compact and point sources to be used for the registration of frames, yielding alignment errors of $<$1 pixel.

\begin{figure}[t!]
\centering
\includegraphics[width=0.48\textwidth]{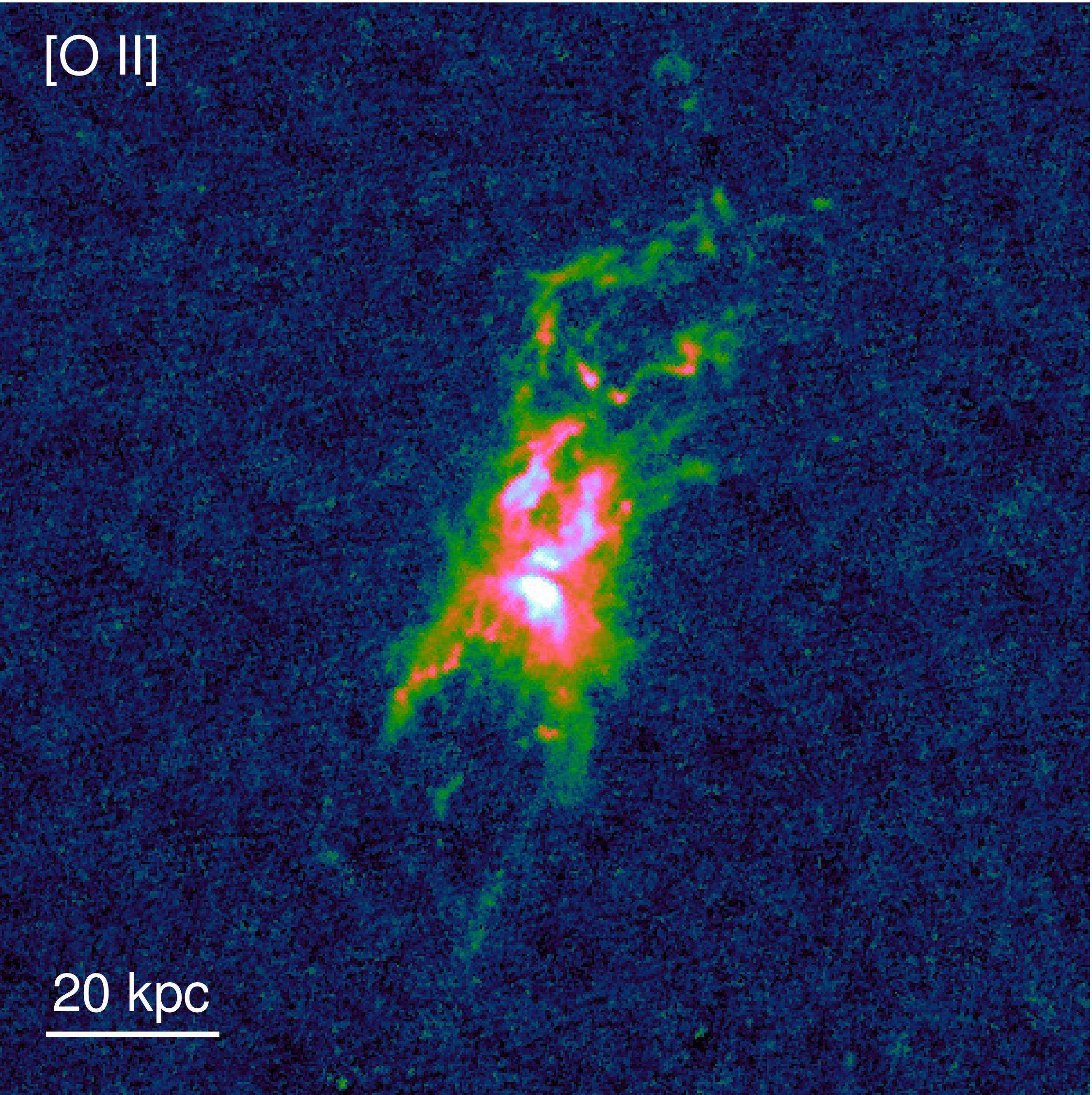}
\caption{Continuum subtracted [O\,\textsc{ii}]$\lambda\lambda$3726,3729 image of the central galaxy in the Phoenix cluster. This image was generated by subtracting a continuum model, based on the three color images in Figure \ref{fig:rgb}, from an image in the narrow-band FR601N filter, as described in \S2.1. This image reveals a significantly more complex network of emission-line filaments than seen in previous ground-based emission line maps \citep{mcdonald14a}. The bulk of the emission is contained in the inner $\sim$20\,kpc, with morphologically-complex filaments extending as far as $\sim$40\, kpc to the south and $\sim$60\,kpc to the north. In general, the emission is confined to a relatively narrow range of position angles, slightly west of north, corresponding to the direction of previously-identified bubbles in the X-ray \citep{hlavacek15}.}
\label{fig:oii}
\end{figure}

To continuum-subtract the narrowband image, we performed a pixel-by-pixel spectral energy distribution (SED) fit. We model the flux from the F475W, F775W, and F850LP filters using the combination of a 10\,Myr old and 6\,Gyr old stellar population, redshifted to $z=0.597$, which were derived from \textsc{Starburst99} \citep{leitherer99}. We allowed for two free parameters: the mass of the young and old stellar populations, and we allowed the normalizations of these components to go negative to preserve noise. This model performed well in subtracting the continuum from a wide variety of populations (as seen in Figure \ref{fig:rgb}), yielding the continuum-free narrowband image shown in Figure \ref{fig:oii}. This image reveals a tremendous amount of structure in [O\,\textsc{ii}] emission that was not seen in previous ground-based narrowband work \citep[e.g.,][]{mcdonald14a}, which we will discuss in \S3.

To calculate the free-fall time ($t_{ff}$) as a function of radius (\S3.2), we estimate the mass profile of the cluster core from its strong lensing signature. A full description of the strong lensing analysis is presented in Bayliss et al. (in prep). In short, we compute a mass model for the cluster core using the public software \lenstool\ \citep{jullo07}, which is a parametric lensing algorithm that uses Markov Chain Monte Carlo to explore the parameter space and identify the best-fit model. We identified nine strongly-lensed sources with multiple images, all but one are spectroscopically confirmed (a full list of constraints is given in Bayliss et al. 2019). The identification of several radial images that extend all the way to the center of the cluster is particularly useful for constraining the mass slope at the cluster core. 

The derivation of a robust lens model and spectroscopic confirmation provide confidence in the identification of the filamentary features near the core, and disentangle emission from background lensed sources from that of the filaments. In particular, this new analysis has allowed us to determine that the 100\,kpc star-forming filament identified in \cite{mcdonald15b}, which contributed $\lesssim$1\% to the total star formation estimate, is actually a lensed source at $z=1.513$. Overall, this model provides a reliable two-dimensional mass constraint over radii of $\sim$1--300\,kpc, which is where we are most interested in constraining the ratio of the cooling to free fall time. This two dimensional model is fit with a projected, three-dimensional model, consisting of a pseudo-isothermal sphere and an NFW profile, to obtain the three dimensional radial mass profile.

\begin{figure*}[t!]
\centering
\includegraphics[width=0.96\textwidth]{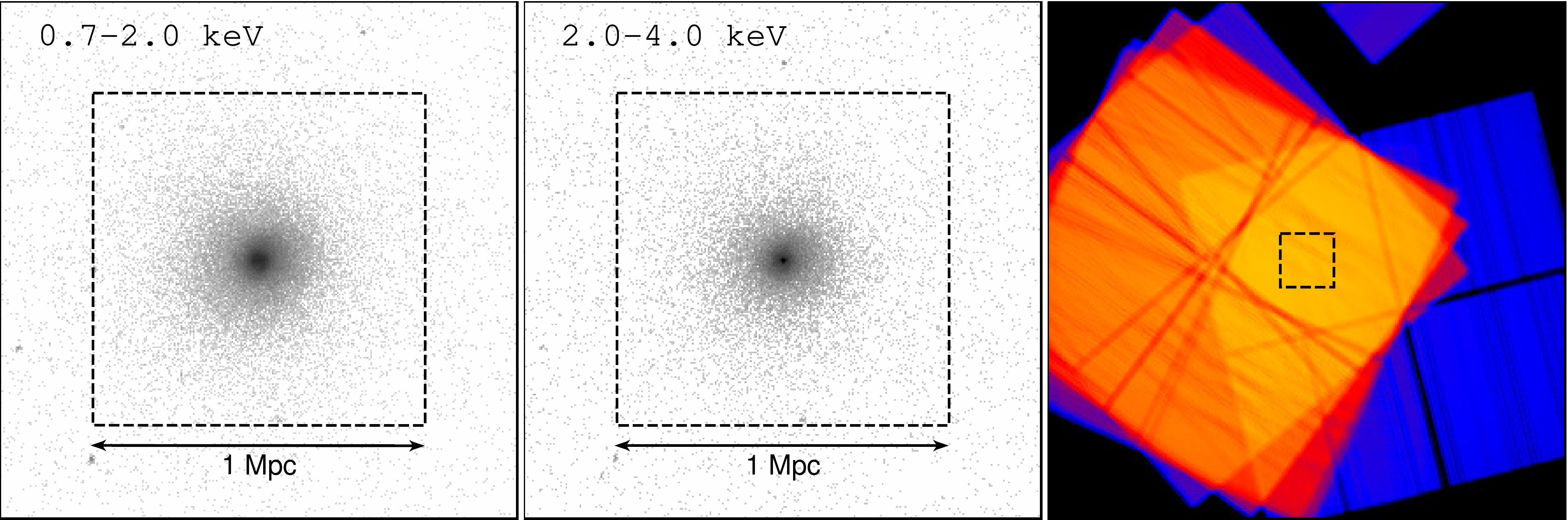}
\caption{\emph{Left:} 0.7--2.0 keV counts image for the inner $\sim$1 Mpc of the Phoenix cluster, combining data from 12 observations totaling 551\,ks. This soft X-ray image shows a relaxed (circular) cluster with a strong cool core. There is no evidence for a central point source at these energies. \emph{Middle:} 2.0--4.0 keV counts image for the same region as shown in the left panel. This hard-band image shows evidence for a central point source, embedded at the center of a strong cool core. The absence of this source in the soft band indicates a highly-obscured AGN. \emph{Right:} Combined exposure map for all 12 observations, with the size of the region from the left panels shown as a small square. This image highlights that all of the relevant data for our analysis (i.e., inner $\sim$1500\,kpc) comes from a small region near the ACIS-I aim point, where we have uniform coverage over all 12 exposure. The bulk of the remaining 3 ACIS-I chips were used as background regions for spectroscopic analyses.}
\label{fig:expmap}
\end{figure*}

\subsection{X-ray: Chandra X-ray Observatory}

X-ray data for the Phoenix cluster was obtained with \emph{Chandra} ACIS-I over a series of programs in Cycle 12 (PI: Garmire, OBSID: 13401), Cycle 15 (PI: McDonald, OBSID: 16135, 16545), and Cycle 18 (PI: McDonald, OBSID: 19581, 19582, 19583, 20630, 20631, 20634, 20635, 20636, 20797). In total, this source was observed for a total of 551\,ks, yielding roughly 300,000 counts in the 0.7--7.0\,keV band (Figure \ref{fig:expmap}). This depth was chosen to allow temperature maps on $\sim$1$^{\prime\prime}$ scales, allowing us to probe structure in the cooling profile on similar scales to the multiphase gas. 

All \emph{Chandra} data were first reprocessed using CIAO v4.10 and CALDB v4.8.0. Point sources were identified on merged images in the 0.7--2.0 and 2.0--7.0 keV bands, using the \textsc{wvdecomp} tool in the \textsc{zhtools} package\footnote{\url{http://hea-www.harvard.edu/RD/zhtools/}}. The resulting list of point sources was used to generate a mask, which was visually inspected. The central QSO in the cluster was removed from the mask. Flares were identified following the procedure outlined by the calibration team\footnote{\url{http://cxc.harvard.edu/ciao/threads/flare/}}, using the 2.3--7.3 keV bandpass, time steps of 519.6s and 259.8s, a threshold of 2.5$\sigma$, and a minimum length of 3 time bins. We used a combination of the ACIS-I stowed background and a local background from the remaining three ACIS-I chips, which were relatively source-free (see Figure \ref{fig:expmap}), following \cite{hickox06}. The stowed background was normalized to match the observations using the measured flux in the 9.0--12.0 keV band.

\subsubsection{Modeling the Central Point Source}

\begin{figure}[t!]
\centering
\includegraphics[width=0.49\textwidth]{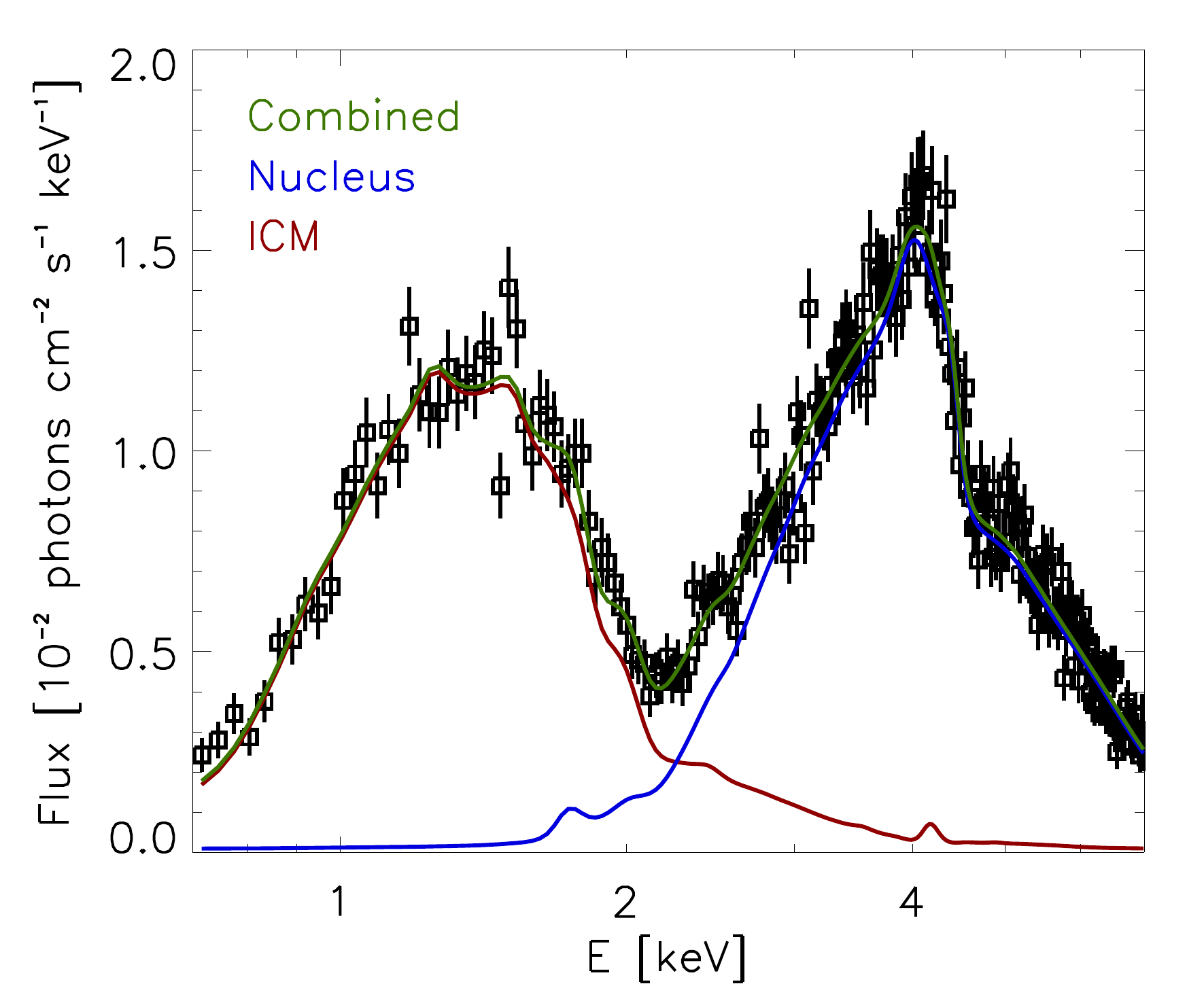}
\caption{Observed-frame X-ray spectrum of the inner 1.5$^{\prime\prime}$ of the Phoenix cluster. This spectrum is dominated at $<$2\,keV by thermal plasma with $kT \sim 2$\,keV, with a small amount of both intrinsic and Galactic absorption (red curve). At $>$2\,keV, the spectrum is dominated by a heavily obscured, marginally piled up central point source, which is well-modeled by the \textsc{MYTorus} model \citep[blue curve;][]{murphy09}. The latter model has been used, in conjunction with \emph{Chandra} ray tracing software, to produce a model of the point source emission for each OBSID, which is used to determine the underlying thermal emission at all radii.}
\label{fig:nucspec}
\end{figure}

The most challenging aspect of the \emph{Chandra} analysis is the modeling of the central point source, a bright type-II QSO \citep{ueda13}, which dominates over the thermal emission in the inner $\sim$10\,kpc, and contributes at similar levels (due to broad PSF wings) to the background over the full area of interest (i.e., $r<1500$\,kpc). 
The X-ray spectrum of the central AGN (inner 1.5$^{\prime\prime}$) is shown in Figure \ref{fig:nucspec}. This double-peaked spectrum explains the relative lack of point source emission at the center of the 0.7--2.0\,keV image (Figure \ref{fig:expmap}).
The spectrum below 2~keV is dominated by the thermal emission from the cluster (modeled with \textsc{apec}), but above 2~keV the AGN emission dominates. The upturn from 2--4~keV and strong iron~K absorption edge is a clear indication a moderately obscured AGN, consistent with the UV to IR SED presented in \cite{mcdonald14a}. The AGN emission in the X-ray spectrum is well described by an absorbed powerlaw with $N_{\mathrm{H}}\sim3 \times 10^{23}$~cm$^{-2}$ ( $\chi^{2}/\mathrm{dof}=466/421=1.11$), however the fit is significantly improved with the addition of an emission line at the rest frame energy of 6.4~keV. Comparing the absorbed powerlaw null hypothesis to one including an additional gaussian fixed at 6.4~keV results in a $\Delta \chi^{2}/\mathrm{dof}=21/3$, equivalent to a $\sim 4\sigma$ detection of a 6.4~keV line.

We fit the spectrum with the \textsc{MYTorus} model, which self-consistently calculates the line-of-sight direct continuum, the scattered emission and fluorescence emission lines from a torus with a $60~\deg$ half-opening angle, surrounding the AGN \citep{murphy09}. \textsc{MYTorus} does not include dynamical effects, and thus we allow for gaussian smoothing of the emission line to model velocity broadening. This model provides a good fit to the data ($\chi^{2}/\mathrm{dof}=448/420=1.07$), and is an improvement over the simple absorbed powerlaw model by $\Delta \chi^{2}/\mathrm{dof}=18/2$ or $\sim 4\sigma$. 
We further include in this model a marginal ($\sim$10\%) amount of pileup, given that the count rate exceeds 0.007 cts/s \citep{davis01}, and intrinsic absorption at the redshift of the Phoenix cluster due to cool gas in the central galaxy. The result of this fit, which has $\chi^2/\mathrm{dof} = 1.02$, is shown in Figure \ref{fig:nucspec}. A more complete interpretation of this spectrum, and of the central QSO in the Phoenix cluster, will be the subject of a future paper.

The best fit model of the central point source (ignoring the \textsc{apec} component which, presumably, comes from a spatially-extended component) was used to produce a simulated ray trace for each observation using ChaRT\footnote{\url{http://cxc.harvard.edu/ciao/PSFs/chart2/}}. This ray trace was combined with MARX\footnote{\url{http://cxc.harvard.edu/ciao/threads/marx/}} to produce a simulated detector-plane image of the central point source for each OBSID. 
When performing spectral fits, we utilize these simulated images to constrain the spectral contribution of the central point source to the total X-ray emission at each position in the detector plane.

\subsubsection{Thermodynamic Profiles and Maps}

We extract profiles in both annuli and two-dimensional regions, with the goal of producing 1-D profiles and 2-D maps of quantities such as temperature ($kT$), electron density, entropy, and pressure. Circular annuli were sampled coarsely at small and large radii, where contributions from the central point source and the diffuse background make measurement of the ICM temperature more challenging, and finely at intermediate radii where the ICM dominates considerably over both the central point source and the background. The spacing of these annuli were chosen iteratively to obtain a uniform uncertainty in the measured temperature over $>$2 orders of magnitude in radius, with this requirement relaxed at the largest radii in the interest of maintaining sampling. Two dimensional regions were generated using the weighted Voronoi tessellation (WVT) binning algorithm of \cite{diehl06}, and were designed to enclose 2000 net counts in the 0.7--2.0 keV bandpass.

Within each region, whether it is a circular annulus or a polygon WVT region, we extract spectra from each OBSID and from each simulated PSF corresponding to the OBSID. In addition, we extract a spectrum for each OBSID of a large off-source area, which is scaled to match the area of the on-source extraction region. We combine these spectra using the CIAO \textsc{combine\_spectra} tool, yielding a high S/N spectrum of the source region, the contribution from the simulated point source, and the nearby background. We joint fit these three spectra with a model that includes: (i) pile-up (\textsc{pileup}), if the region contains any pixels from within 1.5$^{\prime\prime}$ of the central point source,  (ii) Galactic absorption (\textsc{phabs}), (iii) absorption at the redshift of the cluster (\textsc{zphabs}), (iv) an \textsc{APEC} model to account for thermal emission from the ICM, (v) a dusty torus model, as described in the previous section, to model the contribution to the central point source, and (vi) a broken powerlaw with three narrow gaussian lines to model the residual background. The latter model, which is purely phenomenological, provides an excellent fit to the off-source region, with typical residuals at the few percent level, and represents the average residual background above the stowed background, accounting for variations in the overall shape of the background and the line complex at 1--3 keV\footnote{\url{http://cxc.cfa.harvard.edu/contrib/maxim/stowed/}}. When joint-fitting on/off-source data in different annuli, we let only the normalization of the background vary, with all of the shape parameters being held fixed to their value from an independent (background-only) fit.

Spectra are fit in the 0.7--7.0\,keV region, with the upper end chosen to avoid the strong background line at $\sim$7.5\,keV, and the lower end driven by the decreasing effective area and growing uncertainty in the calibration at soft energies due to build up of contamination on ACIS\footnote{\url{http://cxc.harvard.edu/ciao/why/acisqecontamN0010.html}}. Models are tied between on- and off-source (background) regions using the relative area of the extraction regions, and are tied between the data and simulated point source spectra assuming a fudge factor that can vary by $\pm$10\% from unity.  Ultimately, this complex AGN+ICM+background model has only 10 free parameters to fit $\sim$1600 spectral elements (for each spatial element), with the bulk of the spectral shape parameters being fixed to their values from independent fits.

For radial profiles, we convert the measured normalization of the \textsc{apec} model to emission measure ($\int n_e n_H dV$), assuming $N = \frac{10^{-14}}{4\pi [D_A(1+z)]^2}\int n_e n_H dV$, where $D_A$ is the angular diameter distance to the source, $N$ is the normalization of the \textsc{apec} model, $n_e$ is the electron density in units of cm$^{-3}$, and $n_H$ is the hydrogen density in units of cm$^{-3}$. Emission measure profiles were fit by numerically integrating a three-dimensional density profile along the line of sight and over the width of each annulus, producing a projected profile. We assume that the three-dimensional profile is of the form described by \cite{vikhlinin06a}:

{\small
\begin{equation}
n_pn_e = n_0^2 \frac{(r/r_c)^{-\alpha}}{(1+r^2/r_c^2)^{3\beta - \alpha/2}} \frac{1}{(1+r^{\gamma}/r_s^{\gamma})^{\epsilon/\gamma}} + \frac{n_{0,2}^2}{(1+r^2/r_c^2)^{3\beta_2}}~,
\label{eq:dens}
\end{equation}
}

\noindent{}where we leave all parameters free except for $\gamma$, which is fixed to $\gamma=3$, following \cite{vikhlinin06a}. The projected profile is fit to the data using the MPFITFUN procedure in IDL. We fit 100 realizations of the data, where data points are allowed to vary between fits based on their uncertainties, which provides an uncertainty in the fit. To convert from $n_en_p$ to $n_e$, we assume $n_e = \sqrt{n_e^2} = \sqrt{1.199n_en_p}$, where $Z=n_e/n_p=1.199$ is the average nuclear mass for a plasma with 0.3Z$_{\odot}$ metallicity, assuming abundances from \cite{anders89}.

Temperature profiles are similarly projected along the line of sight, assuming a three-dimensional form described by \cite{vikhlinin06a}:

\begin{equation}
T_{3D}(r) = T_0 \frac{(r/r_{core})^{\alpha}+T_{min}/T_0}{(r/r_{core})^{\alpha}+1} \frac{(r/r_t)^{-a}}{[1+(r/r_t)^b]^{c/b}}
\label{eq:temp}
\end{equation}

\noindent{}where the small and large radius behavior is dictated by $r_{core}$ and $r_t$, respectively, and the minimum temperature is $T_{min}$. 
We project this three-dimensional temperature model along the line of sight, and over the width of each bin, using our model density profile from above and assuming that

{\small
\begin{equation}
\left<T\right> = \frac{\int_V wTdV}{\int_V wdV}~,
\label{eq:proj}
\end{equation}
}

\noindent{}where

{\small
\begin{equation}
w=n_e^2T^{-0.75}~,
\end{equation}
}

\noindent{}following \cite{mazzotta04} and \cite{vikhlinin06b}. This projected model was fit to the data, again using MPFITFUN and bootstrapping over 500 realizations of the temperature profile to provide uncertainties on the model. From the three-dimensional temperature and electron density profiles, it is trivial to compute the entropy ($K\equiv kT n_e^{-2/3}$), cooling time ($t_{cool} \equiv \frac{3}{2}
\frac{(n_e+n_p)kT}{n_en_H\Lambda(kT,Z)}$), and pressure ($P=(n_e+n_p)kT$) profiles.

\subsubsection{X-ray Residual Images}

To look for substructure in the X-ray image, we first extract an image in the 0.7--2.0\,keV bandpass and mask any point sources.  We choose this bandpass because it is free of emission from the central AGN, which is almost entirely absorbed at $<$2\,keV, presumably due to cool gas near the central AGN (Figure \ref{fig:nucspec}). This image is modeled in SHERPA, using a combination of two two-dimensional beta models (\textsc{beta2d}). We found that two beta models provided an adequate fit to the emission within 1$^{\prime\prime}$, and that a third component did not significantly improve the fit. We forced the two components to share a center, and to have zero ellipticity, leaving only 6 free parameters: the center ($x$,$y$), the core radii ($r_{c,1}, r_{c,2}$), and the amplitudes. The best-fit model was visually inspected in one dimension to confirm that it was a good fit, and then subtracted from the two-dimensional image to produce a residual image.

\subsection{Radio: Karl Jansky Very Large Array}


The Phoenix Cluster was observed with the Karl G. Jansky Very Large Array (VLA) in the 8--12~GHz X-band (project code 17A-258). The observations were taken in the A-, B-, and C-array configurations on 6 Mar 2018, 5 June 2017, and 7 September 2017, respectively. The total on source time was about 2~hrs in each configuration and all four polarization products were recorded. The primary calibrators used were 3C138 and 3C147. The source J0012-3954 was observed as a secondary calibrator. 

The data was calibrated with CASA \citep{2007ASPC..376..127M} version 5.1. The data for the different array configurations were first reduced separately. The data reduction mostly follows the procedure described in \citet{2016ApJ...817...98V}. Below we briefly summarize the various steps. As a first step the data were corrected for the antenna offset positions, requantizer gains, and elevation dependent gains. Next, data affected by radio frequency interference (RFI) were flagged with the {\tt flagdata} task, employing the `tfcrop' mode. Data affected by antenna shadowing were also removed. We then corrected the data for the global delay, cross-hand delay, bandpass, polarization leakage and angles, and temporal gain variations using the calibrator sources. After averaging the target data, we flagged addition low-level RFI with the {\tt AOFlagger} \citep{2010MNRAS.405..155O}. The calibration solutions were subsequently refined via the process of self-calibration. For the imaging we used WSClean \citep{2014MNRAS.444..606O,2017MNRAS.471..301O} with the Briggs weighting scheme ({\tt robust=0}). Due to the low declination of the target between 10 and 15 cycles of self-calibration were required for the different datasets. The data from the three different array configurations were then combined and jointly imaged. Another 10 rounds  of self-calibration were carried out on the combined datasets. The final images were corrected for the primary beam attenuation.


\section{Results}

\subsection{Cool Gas in the Inner 30\,kpc}

In Figure \ref{fig:rgb}, we see the central starburst in the Phoenix cluster at unprecedented angular resolution. For the first time, the smooth, underlying giant elliptical galaxy can be seen extending on $\sim$50\,kpc scales, with no evidence for tidal disruption as one would expect if the $\sim$600 M$_{\odot}$ yr$^{-1}$ starburst was triggered by a massive gas-rich merger.
This central galaxy is known to have large amounts of gas at 10$^5$\,K \citep{mcdonald15b}, 10$^4$\,K \citep{mcdonald14a}, and in the molecular phase \citep{russell17}. In Figure \ref{fig:oii}, we see the warm ionized (10$^4$\,K) phase, as traced by emission of the [O\,\textsc{ii}]$\lambda\lambda$3726,3729 doublet, at significantly higher angular resolution than any of these previous works. This image reveals a highly clumpy, filamentary complex on similar scales to the emission line nebula in NGC~1275 \citep{conselice01}, spanning $>$100\,kpc ($>$16$^{\prime\prime}$) from end to end. 

 \begin{figure}[t!]
\centering
\includegraphics[width=0.49\textwidth]{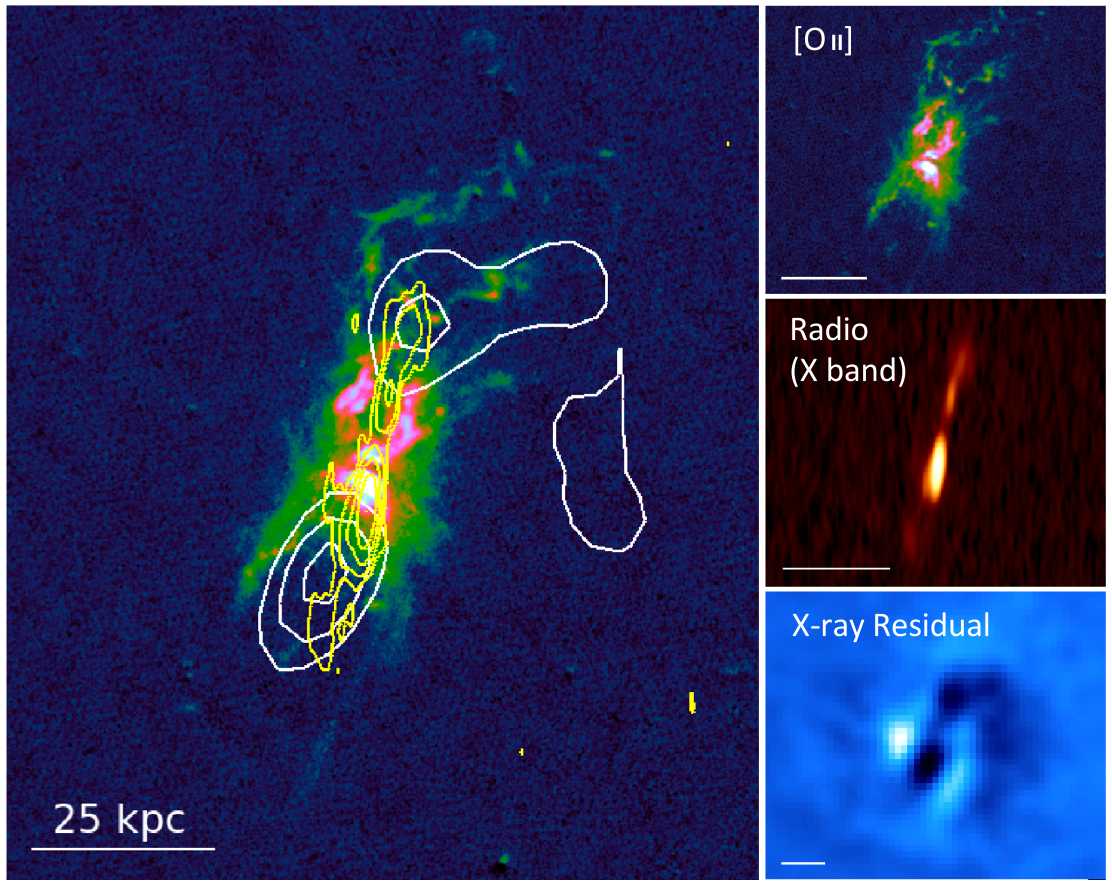}
\caption{On the left, we show the continuum-subtracted [O\,\textsc{ii}] emission map, with radio (yellow) and X-ray (white) contours overlaid. For the X-ray contours, we only show the significant \emph{negative} residuals, after a two-dimensional model has been subtracted from the large-scale emission (see \S2.2.3). In the right panels, we show the [O\,\textsc{ii}] (top), radio (middle), and X-ray (bottom) images individually. X-ray images have been shifted by a half pixel to align the optical and X-ray nuclei. This figure demonstrates the near-perfect correspondence between the radio jets, X-ray cavities, and cool filaments. The jets appear to be responsible for inflating bubbles in the hot gas, while the cool gas appears draped around and behind the resulting cavities. We will discuss this further in \S4.3.}
\label{fig:multiband}
\end{figure}

 \begin{figure}[t!]
\centering
\includegraphics[width=0.48\textwidth]{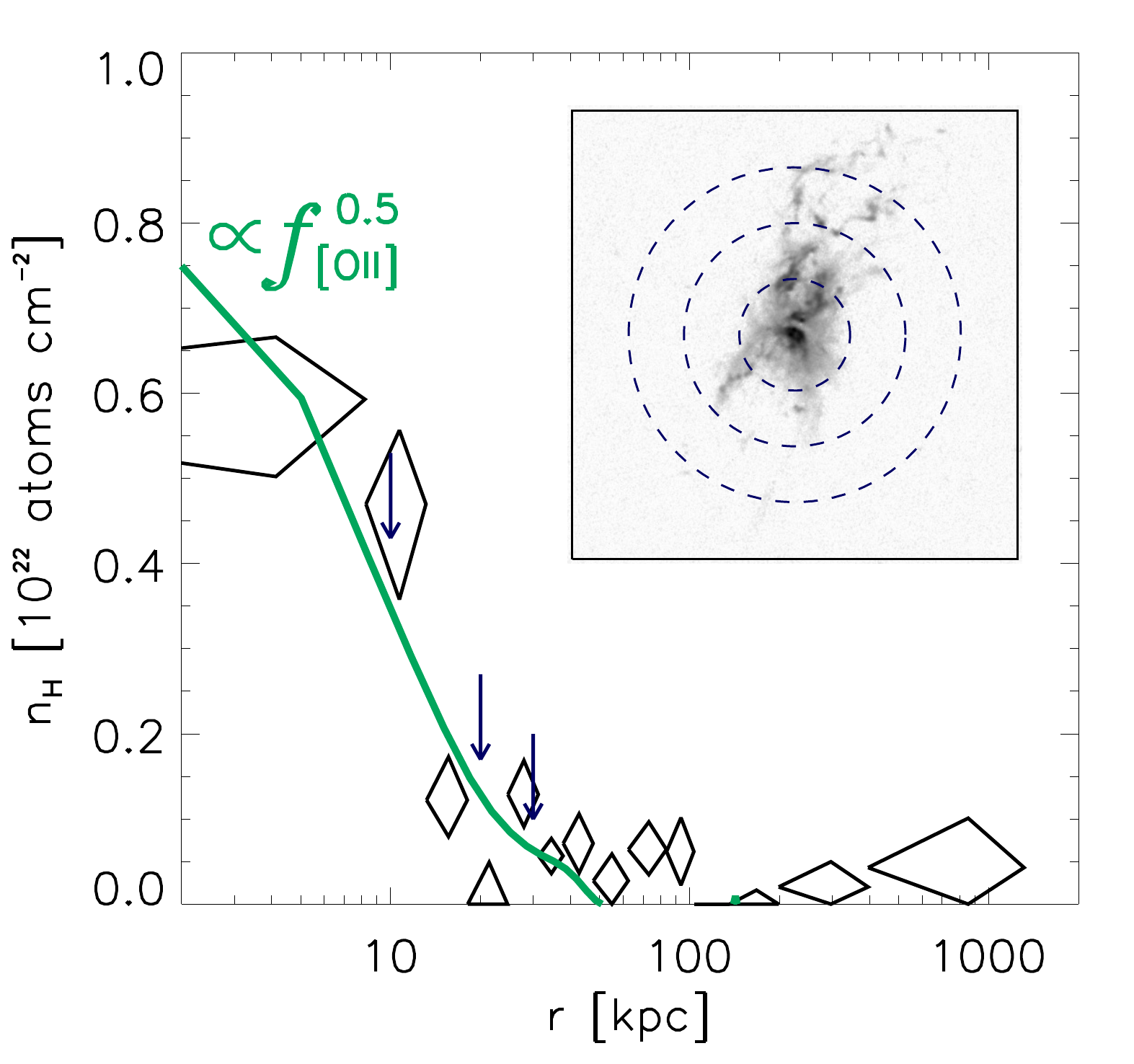}
\caption{This figure shows the inferred photoelectric absorption in the X-ray due to cool gas within the Phoenix cluster as a function of radius. We find relatively strong absorption in the inner $\sim$10\,kpc, which falls off quickly to become negligible at $r\gtrsim50$\,kpc. For comparison, we show the radial profile of the absorbing column (green) as inferred from the [O\,\textsc{ii}] map (shown in grayscale). Assuming that the [O\,\textsc{ii}] emission roughly traces the gas doing the absorbing, then the absorbing column should be proportional to the square root of the emission line luminosity, which qualitatively appears to provide a good fit to the data. Further, integrating the total amount of X-ray absorption over the volume of absorbing material yields a total gas mass similar to that of the cold molecular gas \citep{russell17}. In the inset, we show the [O\,\textsc{ii}] image of the inner $\sim$40\,kpc, with annuli overlaid at 10, 20, and 30\,kpc. These radii are marked in the radial profile with arrows.
}
\label{fig:nHz}
\end{figure}

\begin{figure}[t!]
\centering
\includegraphics[width=0.49\textwidth]{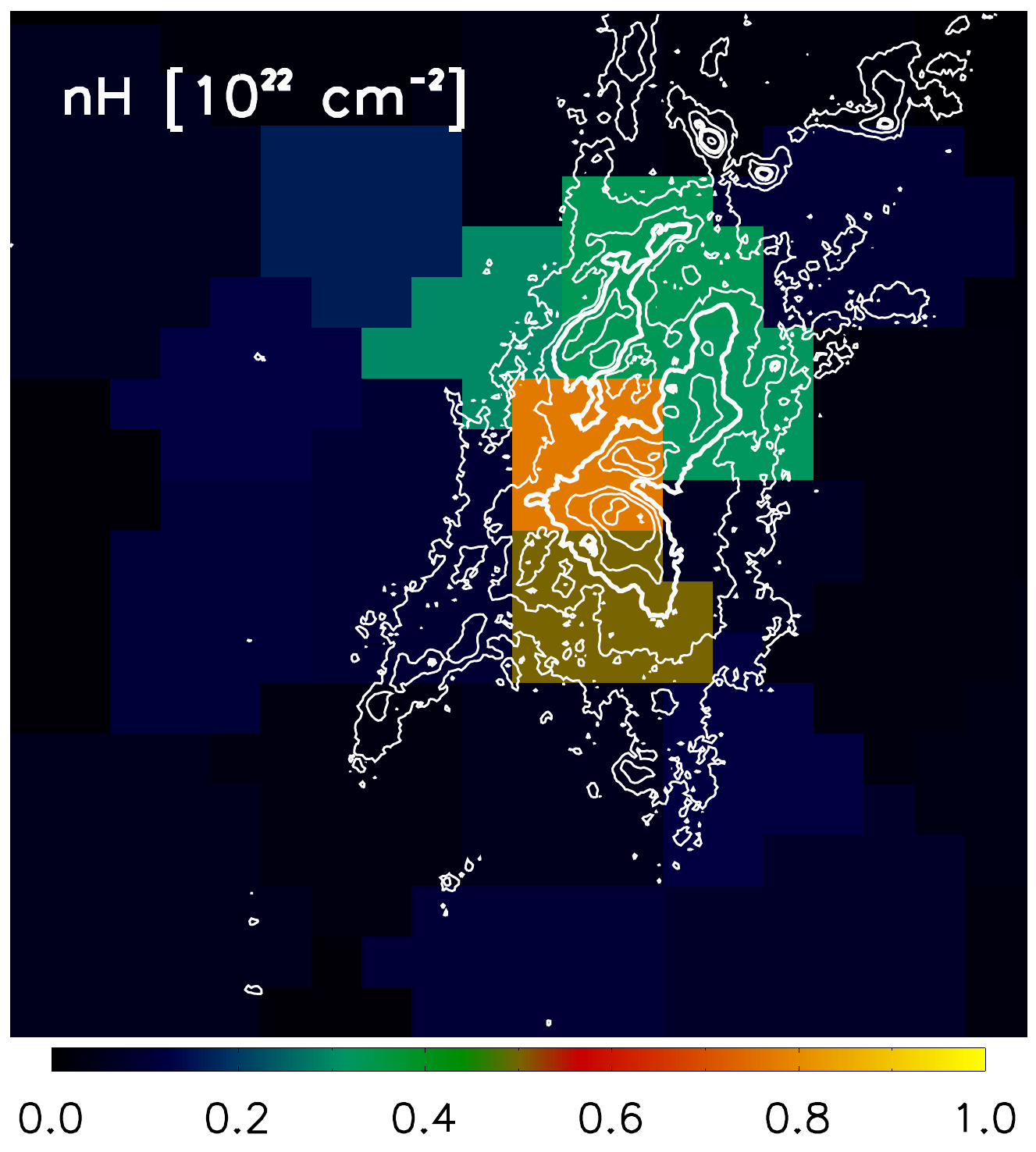}
\caption{Two-dimensional X-ray absorption map for the inner $\sim$25\,kpc of the Phoenix cluster. Overlaid in white are the contours of constant [O\,\textsc{ii}] surface brightness, where we have highlighted the highest surface brightness emission with a thicker contour. This figure demonstrates that the X-ray-absorbing material is cospatial with the cool, line-emitting gas. 
}
\vspace{-0.1in}
\label{fig:nHzmap}
\end{figure}

When we compare this map to both the residual X-ray map and the high resolution radio map, we see several morphological similarities between the cool gas and the radio jets. To the north and south, the cool gas appears draped around the outside of the bubbles, similar to what is observed in CO \citep{russell17}. There are several filaments that extend beyond the bubbles, including a linear filament extending $\sim$10--20\,kpc below the southern bubble and a curved northern filament extending $\sim$30\,kpc beyond the leading edge of the northern bubble. These filaments challenge the notion that the cool gas is condensing exclusively behind bubbles. The bulk of the [O\,\textsc{ii}] emission lies between the location of the central AGN and the southern edge of the northern bubble. This cool gas appears similar to the ``horseshoe'' filament in Perseus, which may be gas condensing behind the bubble as it uplifts low-entropy gas in its wake.

Given the huge amount of cool gas in this system, and the high covering fraction over the inner $\sim$30\,kpc, one would expect a significant amount of absorption in the soft X-rays. Indeed, if we look at the best-fit intrinsic absorption column as a function of radius (Figure \ref{fig:nHz}), we see that it is significantly elevated in the inner $\sim$30\,kpc. We note that the inclusion of intrinsic absorption absorption improves the quality of fit from $\chi^2_{dof} = 1.39$ for a single temperature model or $\chi^2_{dof} = 1.42$ for a multi-temperature model to $\chi^2_{dof} = 1.22$ for an absorbed single temperature model.
The absorption profile is proportional to the square root of the [O\,\textsc{ii}] flux profile, as one would expect if the cool gas is the primary source of absorption. Further, if we consider the two-dimensional X-ray absorption map (Figure \ref{fig:nHzmap}), we find that the morphology is strikingly similar to the [O\,\textsc{ii}] map, extending primarily to the north and peaking at the galaxy center. 
Assuming that this absorbing material is distributed evenly throughout a cylinder 40\,kpc long and 20\,kpc in diameter, and assuming an average column of $0.5\times10^{22}$ cm$^{-2}$ (Figure \ref{fig:nHzmap}), we infer a total mass of $1.2\times10^{10}$ M$_{\odot}$ in cool material, which is comparable to what we see in the molecular phase \cite{russell17}. Assuming a cooling rate of $\sim$1000\,M$_{\odot}$ yr$^{-1}$, this amount of cool material could accumulate in only $\sim$10$^7$ years.

Such substantial absorption has not been detected in other more nearby clusters. This can be understood by considering a direct comparison to the Perseus cluster, which is the archetypal cool core cluster. 
First, the Phoenix cluster is in a part of the sky nearly free of Galactic absorption, with a Galactic absorbing column of $0.015 \times 10^{22}$ cm$^{-2}$, compared to $0.14 \times 10^{22}$ cm$^{-2}$ for the Perseus cluster. Second, the Phoenix cluster is at a redshift of $z=0.597$, compared to Perseus which is at $z=0.018$. This means that any intrinsic absorption in Phoenix will affect the spectral shape differently than Galactic absorption, unlike in Perseus where both intrinsic and Galactic absorption have essentially the same spectral signature, with only an amplitude offset. Finally, the cool gas in the Phoenix cluster appears to be less filamentary than in the Perseus cluster, which may lead to a higher covering fraction of the absorbing gas.
Combining these three factors, we expect for Phoenix the ratio of intrinsic to Galactic absorption to be a factor of $\sim$100 times higher than for the Perseus cluster, while also having a different spectral signature than Galactic absorption, making it easier to detect.

Overall, we see evidence for an abundance of cool gas based on both high angular resolution [O\,\textsc{ii}] imaging and low angular resolution X-ray absorption mapping. The [O\,\textsc{ii}] morphology is well-matched by the X-ray absorption morphology, both of which appear to be highly concentrated to the north of the central galaxy, in the direction of the radio jet and behind the northern X-ray cavity. We will discuss the implications of this large cool gas reservoir in \S4.

\subsection{Thermodynamic Profiles: Evidence for a Cooling Flow}

\begin{figure}[t!]
\centering
\includegraphics[width=0.48\textwidth]{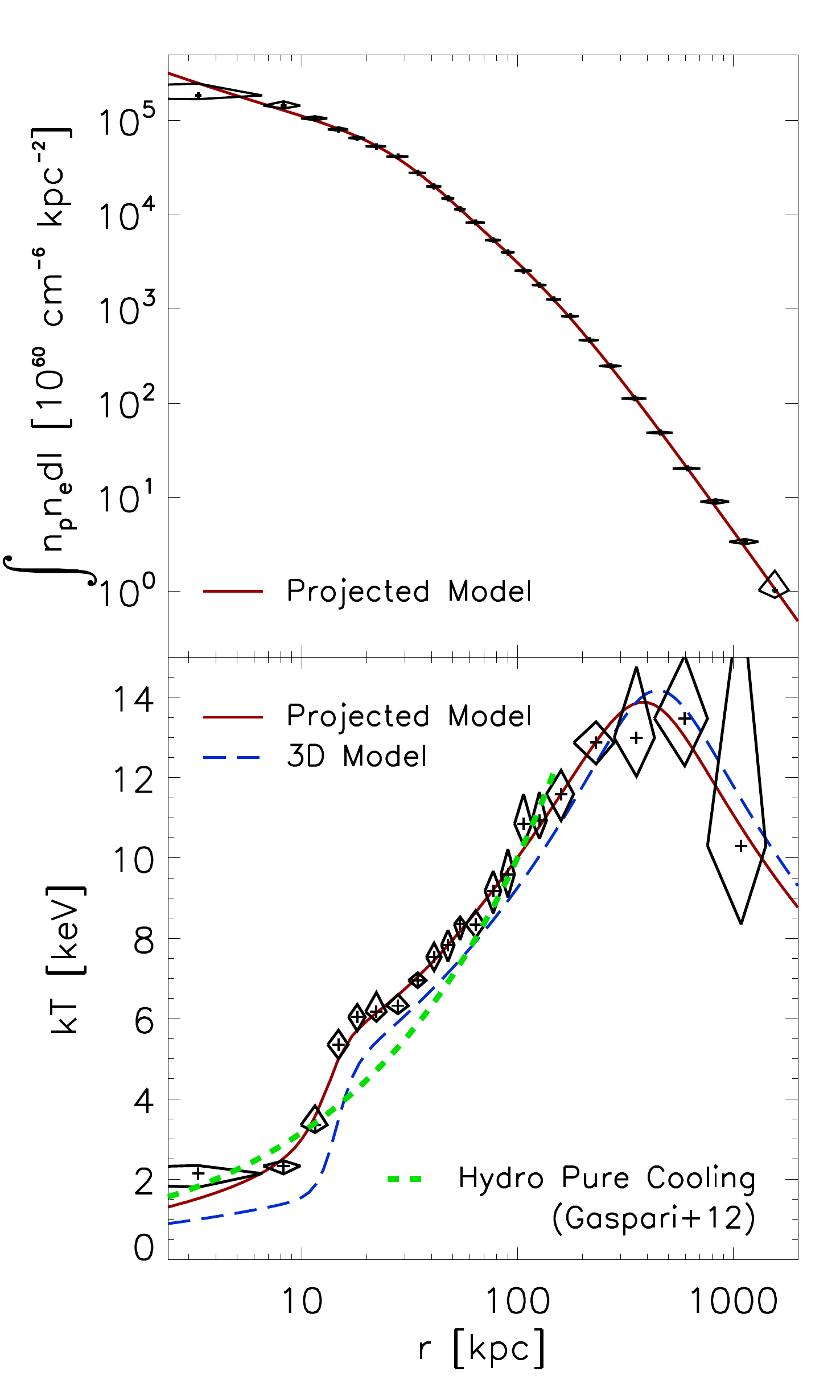}
\caption{\emph{Top:} Projected X-ray emissivity per unit area for the Phoenix cluster. In red, we show the best-fit three-dimensional model (Equation \ref{eq:dens}), projected onto two dimensions. This model provides an excellent fit to the data at all radii and reveals a strongly-peaked cool core cluster. \emph{Bottom:} Projected X-ray temperature profile for the Phoenix cluster. In red, we show the best-fit three-dimensional model (Equation \ref{eq:temp}), projected onto two dimensions. In blue we show the true three-dimensional model, which is slightly cooler in the center and slightly hotter at large radii than the projected profile. In green, we show the prediction for the inner 150\,kpc from hydrodynamic simulations for a pure cooling flow \citep[$T \propto r^{1/2}$;][]{gaspari12}. Phoenix exhibits the strongest fall-off in temperature of any cluster known, going from a peak temperature of $\sim$14\,keV to a minimum of $\sim$1\,keV in the two innermost bins ($r \lesssim 10$\,kpc), consistent with the expectations for pure cooling. The temperature profile appears to have two regions: the larger scale cool core with a temperature of $\sim$6\,keV and a size of $\sim$50\,kpc, and a smaller, cooler region with a temperature of $\sim$1\,keV and a size of $\sim$10\,kpc.
}
\label{fig:xray_profs}
\end{figure}

In Figure \ref{fig:xray_profs}, we show the one-dimensional emission measure and temperature profiles for the Phoenix cluster. The emission measure profile is well fit by our projected density model at all radii, and looks similar to the majority of cool cores, with the density steadily rising towards the cluster center. In the central 10\,kpc, the three dimensional electron density surpasses 0.5 cm$^{-3}$, which is more typical of the warm neutral/ionized medium of a disk galaxy, and roughly an order of magnitude higher than the typical cool core cluster at $z\sim0$ \citep{sanderson09,hudson10}. The projected temperature profile decreases from a peak of $\sim$14\,keV at $r\sim300$\,kpc to a minimum of $\sim$2\,keV in the center, while the three-dimensional profile indicates that the ambient central temperature may be as low as $\sim$1\,keV.. This central temperature corresponds to only $\sim$1\,keV in three dimensions. Over $\sim$300\,kpc in radius, this represents the strongest temperature gradient in any known cool core cluster, and is consistent with the $T \propto r^{1/2}$ expectation for pure cooling in hydrodynamic simulations \citep{gaspari12}. In the inner $\sim$20\,kpc, the temperature appears to fall off even more rapidly, dropping from 6\,keV to 1\,keV over $\sim$10\,kpc. The lack of an accompanying density jump suggests that this is not a cold front.

\begin{figure}[t!]
\centering
\includegraphics[width=0.48\textwidth]{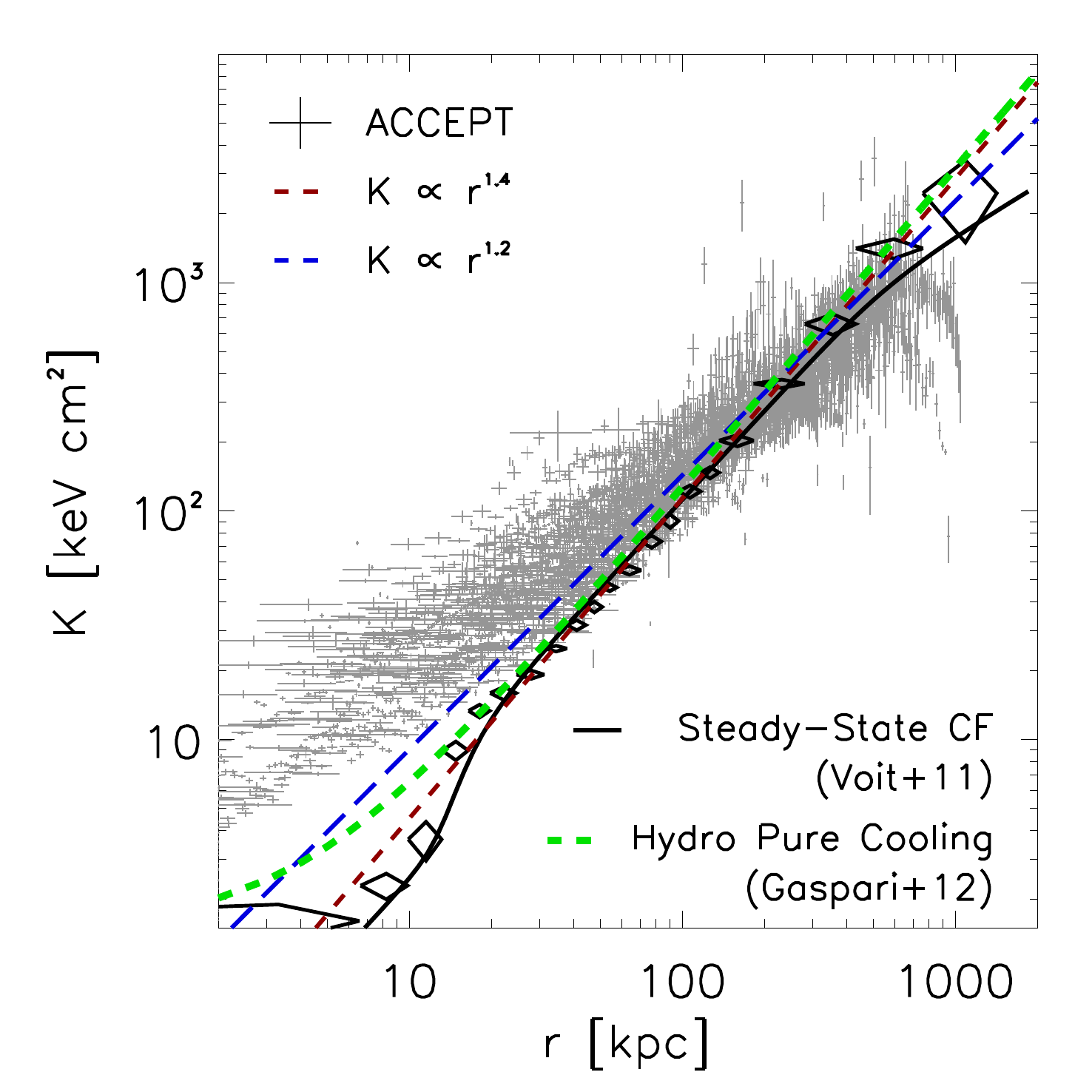}
\caption{Three-dimensional entropy profile for the Phoenix cluster (black diamonds). This profile was derived based on the best-fit three-dimensional density and temperature models, shown in Figure \ref{fig:xray_profs}. In gray, we show entropy profiles for 91 cool core clusters ($K_0 < 30$\,keV cm$^2$) in the ACCEPT sample \citep{cavagnolo09}. In blue we show the canonical $K\propto r^{1.2}$ ``baseline entropy profile'' from \cite{voit05}, while in red and black we show the simple and more complicated (respectively) predictions for a steady-state cooling flow with \.M $\sim$ 3000 M$_{\odot}$ yr$^{-1}$, following \citep{voit11}. In green, we show the prediction for a pure cooling flow from hydrodynamic simulations \citep{gaspari12}. We note that, at all radii, the data for the Phoenix cluster lies between the simulation and analytic theory predictions for a pure cooling flow. This plot demonstrates the uniqueness of the Phoenix cluster, which exhibits no evidence for a break at small radii \citep[e.g.,][]{panagoulia14,babyk18} and with a central entropy lying roughly an order of magnitude below any other cluster known. }
\label{fig:kprof}
\end{figure}

In Figure \ref{fig:kprof} we show the three dimensional entropy profile for the Phoenix cluster, compared to cool cores from the ACCEPT sample \citep{cavagnolo09}. At large radii ($r>100$\,kpc), the entropy profile appears to be slightly steeper than the baseline $r^{1.2}$ profile produced by gravitational structure formation \citep{voit05}, but is consistent in magnitude with the bulk of nearby cool core clusters. This steep slope continues over $30 < r < 100$\,kpc, at which point the entropy profile for most other clusters has become shallower \citep[e.g.,][]{panagoulia14,babyk18}. At a radius of 30\,kpc, Phoenix has the lowest entropy of any nearby cluster, and appears to be following a single power with slope r$^{1.4}$. 
At $r<30$\,kpc, specific entropy continues to decline toward smaller radii, steepening at $10 < r < 20$\,kpc due to the sharp drop in the temperature profile (Figure \ref{fig:xray_profs}). In the innermost bin, $0 < r < 7$\,kpc, we measure a three-dimensional entropy of $\sim$2\,keV cm$^2$, roughly 5$\times$ lower than any other cluster at similar radii in the ACCEPT sample. Over the range $0 < r < 30$\,kpc, the entropy profile lies below the baseline entropy profile, as one would expect for a rapidly-cooling system.

We also show in Figure \ref{fig:kprof} the expectation for pure cooling in hydrodynamic simulations \citep{gaspari12}, as well as an analytic steady-state cooling flow model, as described by \cite{voit11}. In this model, a steady state entropy profile is achieved when inward advection of heat balances radiative losses:

\begin{equation}
\frac{P}{(\gamma-1)}\frac{v}{r}\frac{d\rm{ln}K}{d\rm{ln}r} = -\rho^2\Lambda(T)
\label{eq:voit1}
\end{equation}

\noindent{}where $v$ is the inflow speed and $\gamma$ is the polytropic index. Assuming $\gamma=5/3$, $\Lambda(T)=1.7\times10^{-27}m_p^{-2}T^{1/2} ~\rm{erg}~\rm{cm}^3~\rm{s}^{-1}~\rm{K}^{-1/2}$, and recognizing that \cite{voit11} define entropy with respect to mass density ($K_{V11}=P\rho^{-\gamma} = K/(\mu\mu_e^{2/3}m_P^{5/3}$), yields:

\begin{equation}
K(r) = \frac{3.9 ~\rm{keV} \rm{cm}^2}{(\dot{\rm{M}}/\rm{M}_{\odot}~\rm{yr}^{-1})^{1/3}} \times \left[ \int_0^{r_{kpc}} \left(\frac{kT}{\rm{keV}}\right)^{5/2} r_{kpc}^2dr_{kpc}\right]^{1/3}
\label{eq:voit2}
\end{equation}

\noindent{}where we use the more standard X-ray definition of $K\equiv kT n_e^{-2/3}$, have assumed $\mu=0.6$ and $\mu_e=1.165$, and have explicitly declared the units for each variable for clarity.
 \cite{voit11} shows that, for most clusters, this expression results in a power law profile with $K \propto r^{1.2}$. However, this slope is weakly dependent on the slope of the temperature profile. Given the very steep temperature profile observed in Phoenix ($kT \propto r^{0.5}$), the implied steady-state entropy profile has $K \propto r^{1.4}$, which we have overplotted on Figure \ref{fig:kprof}.  The precise expectation for the entropy profile in a steady-state cooling flow, derived from Equation \ref{eq:voit2} and the three-dimensional temperature profile, is shown as a solid black line in Figure \ref{fig:kprof}. Renormalizing this model to match the data implies a cooling rate of 3276 M$_{\odot}$ yr$^{-1}$ and an inflow velocity ($v_{flow} \equiv \dot{M}/4\pi r^2 \rho$) of $<$200\,km s$^{-1}$ at $r>10$\,kpc and $<$50\,km s$^{-1}$ at $r>50$\,kpc. This inflow speed would become supersonic at $r\sim5$\,kpc -- these scales, which correspond to one ACIS pixel, are beyond our ability to resolve given the bright central point source.

\begin{figure}[t!]
\centering
\includegraphics[width=0.49\textwidth, trim={0 0 19.5cm 0 }, clip]{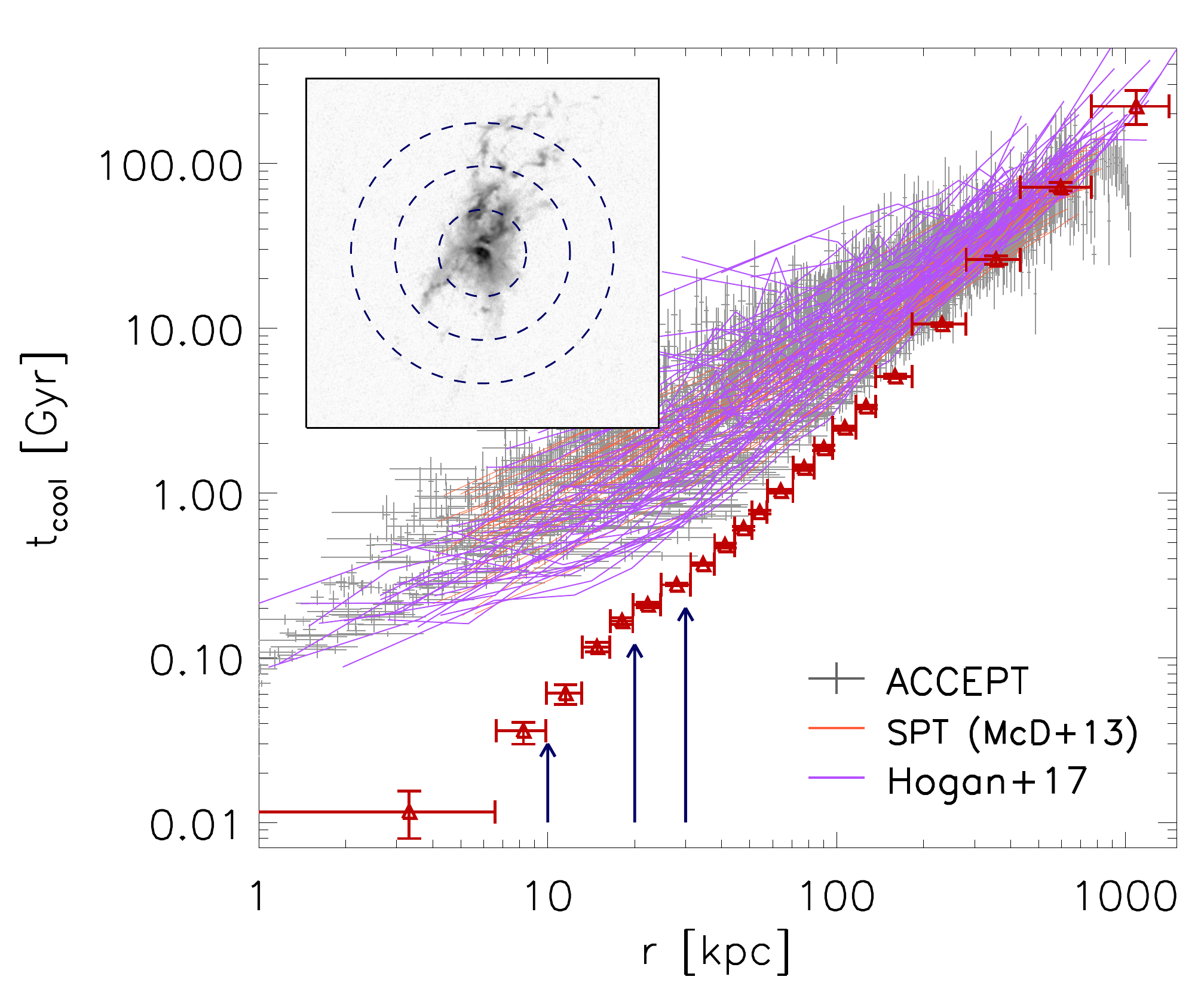}
\caption{Cooling time as a function of radius for the Phoenix cluster (red points). We show, for comparison, 91 cool core clusters in the ACCEPT sample \citep{cavagnolo09}, 57 cool core clusters from the sample of \cite{hogan17}, and 27 high-$z$ cool core clusters from \cite{mcdonald13b}. In the inset, we show the [O\,\textsc{ii}] image of the central galaxy, with radii of 10\,kpc, 20\,kpc, and 30\,kpc depected as dashed circles. These radii are also highlighted in the profile, showing that the bulk of the cool gas is present at radii where the cooling time is $\lesssim$100\,Myr. At these radii, the Phoenix cluster has the shortest cooling time of any known cluster by a substantial margin.}
\label{fig:tcool}
\end{figure}

\begin{figure}[t!]
\centering
\includegraphics[width=0.49\textwidth]{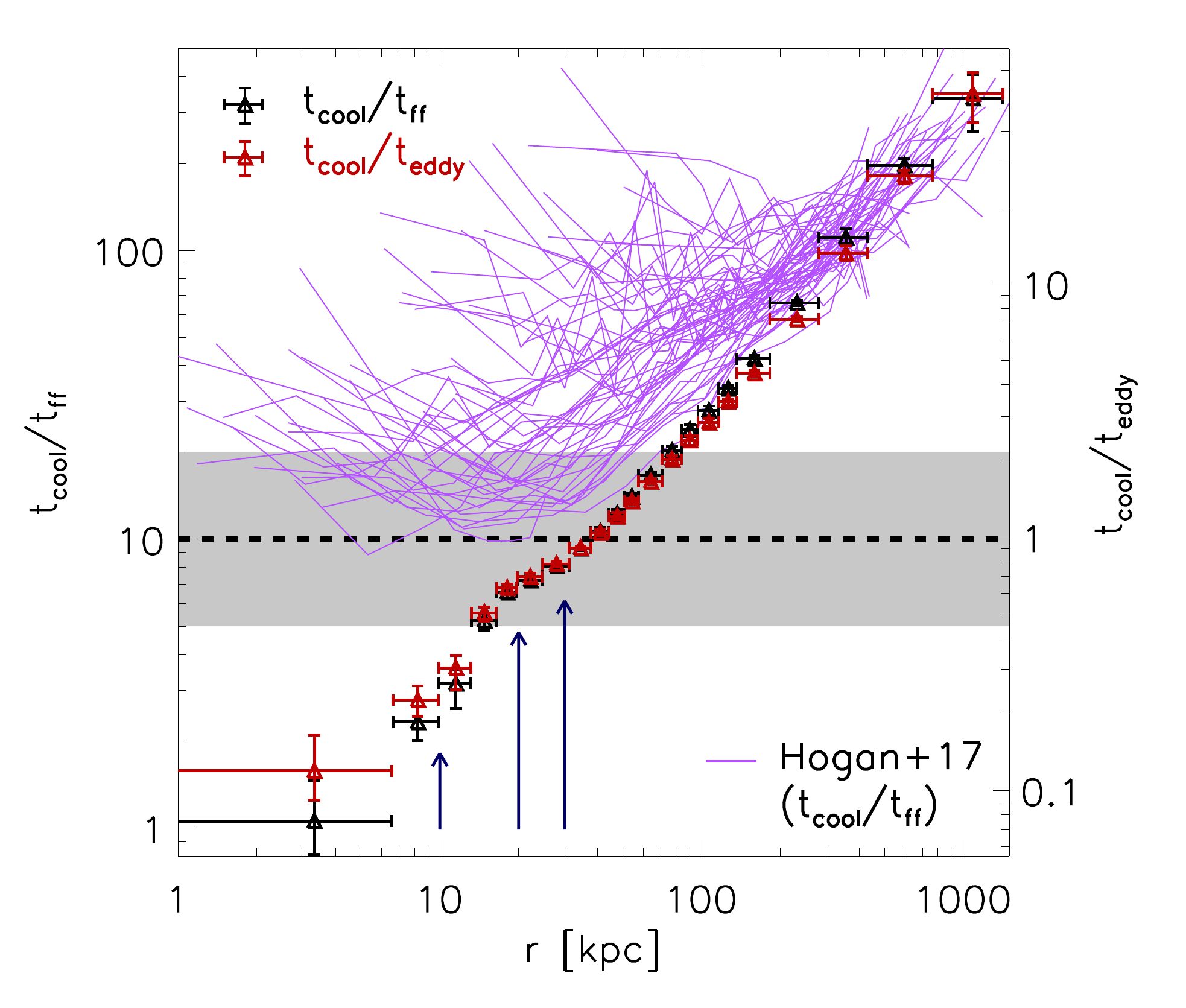}
\caption{Cooling time normalized to the free fall time ($t_{cool}/t_{ff}$; black) and to the eddy time ($t_{cool}/t_{eddy}$; red)  as a function of radius for the Phoenix cluster. These ratios have been used as predictors for the growth of thermal instabilities \citep[e.g.,][]{gaspari12,mccourt12,sharma12,voit15,gaspari18}, where the denominator is meant to roughly approximate the mixing time of the gas. Here, the freefall time has been calculated using the strong lensing model model, described in \S2.1, while the eddy timescale has been computed assuming a line-of-sight turbulence of 250 km s$^{-1}$ and a bubble diameter of 20\,kpc \citep[see Eq.\ 5 in][]{gaspari18} . We show, for comparison, 57 clusters from \cite{hogan17}, for which the $t_{cool}/t_{ff}$ ratio has been computed with a careful consideration of the gravitational potential due to both the cluster and the central galaxy. The Phoenix cluster is an obvious outlier in $t_{cool}/t_{ff}$ space, as the only known cluster reaching a value of unity in the core. The dotted black line and shaded grey region highlights the condensation threshold for these two ratios. 
This figure demonstrates that thermal instabilities are likely to develop within the central 30 kpc, where we see a significant amount of cool gas.}
\label{fig:tctff}
\end{figure}

In Figure \ref{fig:tcool} we show the cooling time ($t_{cool}$) profile for the Phoenix cluster. In the inner $\sim$15\,kpc of the Phoenix cluster, the cooling time falls below 100\,Myr, reaching as low as $\sim$10\,Myr in the innermost bin. At 10\,kpc, the cooling time is $\sim$50\,Myr, shorter by a factor of $\sim$5 than any of the 91 cool core clusters in the ACCEPT sample \citep{cavagnolo09}, any of the 27 high-$z$ cool core clusters from the SPT-\emph{Chandra} sample \citep{mcdonald13b}, or any of the 56 nearby, well-studied cool core clusters from \cite{hogan17}. It is clear from this plot, which includes nearly every known cool core cluster spanning $0 < z < 1.5$, that the Phoenix cluster is an outlier, and likely subject to a different evolution than the typical cluster.

We show in Figure \ref{fig:tctff} the ratio of the cooling time to the freefall time ($t_{cool}/t_{ff}$) where the freefall time has been inferred from the strong lensing mass profile (\S2.1). A value of $t_{cool}/t_{ff} = 10$ has been proposed as a threshold 
below which ambient gas becomes highly susceptible to multiphase condensation \citep{gaspari12,mccourt12,sharma12,voit15}, and \cite{voit18a} has shown that, over 3 orders of magnitude in mass, spanning galaxies like our Milky Way up to the richest galaxy clusters, no known system has $t_{cool}/t_{ff} < 5$ in its hot, diffuse halo. We find no evidence for a minimum in the $t_{cool}/t_{ff}$ profile, which reaches $\sim$1 in the innermost bin. This is an order of magnitude lower than in any other known group or cluster \citep{hogan17}, and the lowest value measured in any diffuse X-ray halo \citep{voit18a} with the exception of the inner 100\,pc of M87 \citep{russell18}.  The fact that Phoenix is the one system that we know of where the hot, diffuse gas reaches $t_{cool}/t_{ff} \sim 1$ on kpc scales, while simultaneously hosting the most star-forming central cluster galaxy, is likely not a coincidence. 

For comparison, we also show in Figure \ref{fig:tctff} the ratio of the cooling time to the eddy time ($t_{cool}/t_{eddy}$), which has been proposed by \cite{gaspari18} as an alternative timescale to govern the condensation of hot gas. Here, the eddy time has been computed following \cite{gaspari18} and assuming $\sigma_{los} = 250$ km/s for the warm gas \citep{mcdonald14a} and a bubble diameter of 20\,kpc \citep{hlavacek15}.  \cite{gaspari18} propose that thermal instabilities will develop when $t_{cool}/t_{eddy} < 1$, which is typically satisfied at radii of $\sim$10--20 for galaxy clusters. The Phoenix cluster is not unique in terms of its $t_{cool}/t_{eddy}$ profile, which is surprising given how big of an outlier it is in most other plots.
Interestingly, the radial dependence of the eddy timescale is very similar to the free-fall timescale in the Phoenix cluster, which is not the case for most other clusters \citep{gaspari18}. As such, we find that $t_{cool}/t_{ff} < 10$ and $t_{cool}/t_{eddy} < 1$ at $r\lesssim30$\,kpc, with the two profiles being statistically indistinguishable in Figure \ref{fig:tctff}. This denominator-independent behavior is primarily driven by the steepness of the cooling time profile, which has $t_{cool} \propto r^{1.4}$, compared to the much shallower eddy timescale ($\propto r^{0.67}$) or freefall timescale ($\propto r^{0.8}$).

Both the entropy and cooling time profiles exhibit behaviors that are unmatched in any other known cluster. The single-powerlaw nature of these profiles, coupled with the fact that the $t_{cool}/t_{ff}$ ratio reaches a minimum of $\sim$1, is consistent with a 
steady, homogeneous cooling flow that is becoming thermally unstable in the inner $\sim$10--20\,kpc. We will return to this idea in more detail in the discussion below.

\subsection{Thermodynamic Maps -- Evidence for Asymmetric Cooling}
  
 \begin{figure*}[t!]
\centering
\includegraphics[width=0.95\textwidth]{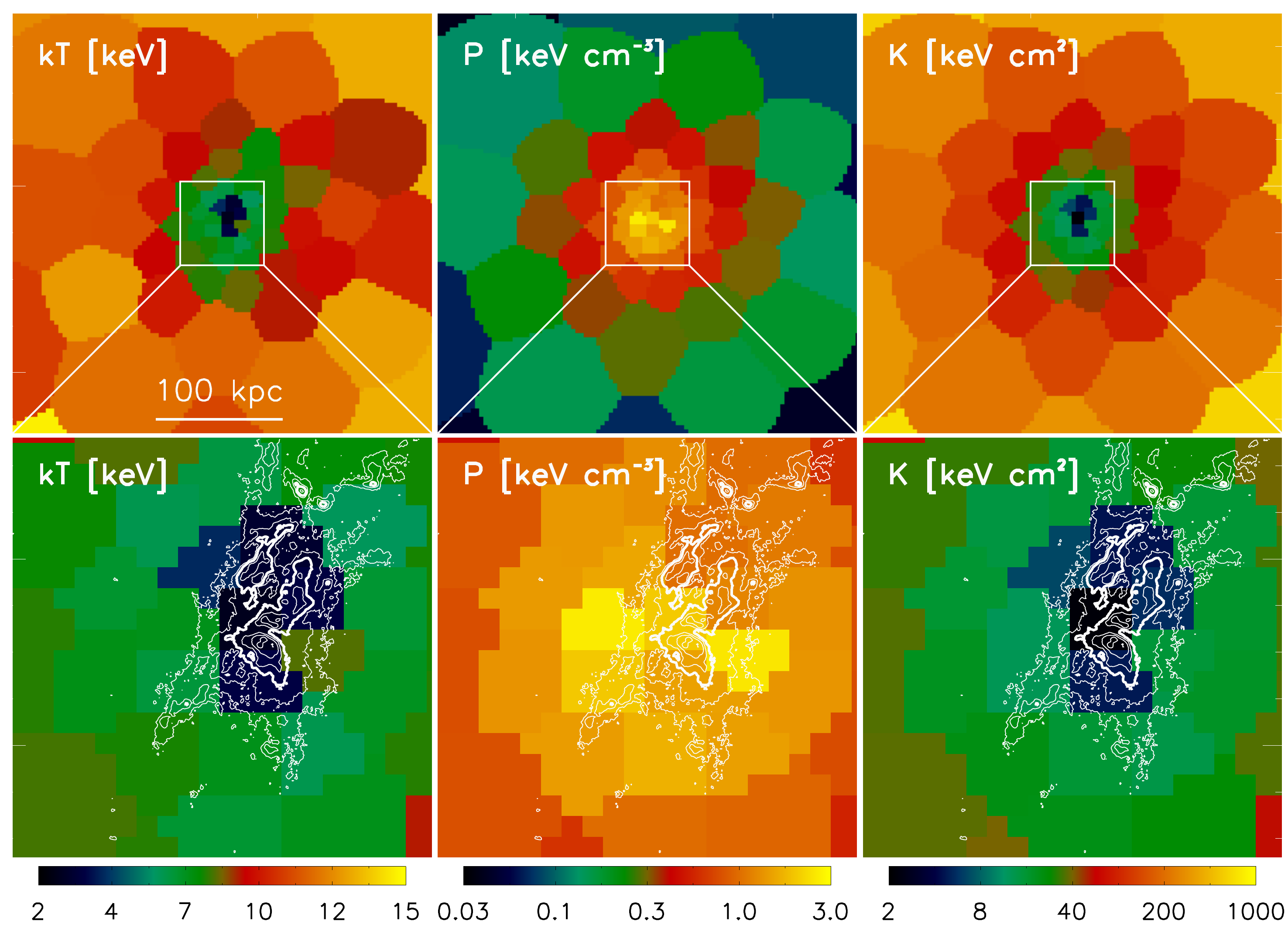}
\caption{Thermodynamic maps, showing temperature (left), pressure (middle), and entropy (right). Images are $330\times330$\,kpc (top) and $66\times66$\,kpc (bottom) on a side. In the lower panels, we show contours of constant [O\,\textsc{ii}] surface brightness, highlighting the location of the cool, line-emitting gas. We highlight the highest surface brightness emission with a thick white contour. Both the temperature and entropy profiles show a filament of cool, low-entropy gas to the north, extending $\sim$20 kpc. This asymmetric low-entropy region, which is also the location of the strongest X-ray absorption (see Figure \ref{fig:nHzmap}) is cospatial with the cool, line-emitting gas. This region appears to be in, or close to, pressure equilibrium with the surrounding regions. There is some weak evidence for enhanced pressure to the east and west, which is primarily driven by an increase in density (i.e., the accompanying temperature increase is mild). 
}
\label{fig:specmaps}
\end{figure*}
  
Thus far, the thermodynamic analysis has been purely 1-dimensional, assuming circular symmetry -- here, we investigate the possibility that cooling may be asymmetric. In Figure \ref{fig:specmaps}, we show the 2-dimensional temperature, pressure, and entropy maps for the inner $\sim$100\,kpc of the Phoenix cluster. These maps are made using the weighted Voronoi tessellation binning algorithm, as described in \S2.2.2. The temperature, pressure, and entropy maps show a relatively symmetric, circular cool core on large scales. On small scales, the coolest gas appears to be concentrated along a filament extending $\sim$20\,kpc to the north, reminiscent of the cool core in Abell~1795 which has a similar X-ray filament extending to the south of the central galaxy \citep[e.g.,][]{crawford05,mcdonald09,ehlert15}. This northern, low-entropy filament is cospatial with both the northern radio jet, and the bulk of the cool gas, as shown in Figures \ref{fig:multiband}, \ref{fig:nHzmap}, and \ref{fig:specmaps}. The spatial resolution of the X-ray data, combined with the difficulty of modeling the central point source, precludes us from making thermodynamic maps on similar scales to the HST or VLA data. To the degree that we can constrain its spatial distribution, the lowest entropy gas, with $K_{2D} \sim 5$\,keV cm$^2$, appears to be confined to the same region as the highest column density [O\,\textsc{ii}], suggesting that these two phases are linked. In the pressure map, there is weak evidence ($\sim$2$\sigma$) that the gas is either under-pressured in the north-south direction, or over-pressured in the east-west direction, or both. 
If gas is condensing along the cooling filament faster than a sound wave can cross it \citep[e.g.,][]{gaspari15}, it ought to lead to an under-pressured region such as this. 
Given that the cooling time of the gas in this filament is $\sim$10\,Myr, and the sound-crossing time of a 10\,kpc wide cooling filament is $t_{cross} \sim l/c_s \sim 10$\.kpc$ / [1480$ km/s $(T_g/10^8K)^{1/2}] \sim 20$\,Myr, one may expect the gas to be slightly under-pressured in the region where gas is condensing most rapidly.
%

\begin{figure}[h!]
\centering
\includegraphics[width=0.49\textwidth]{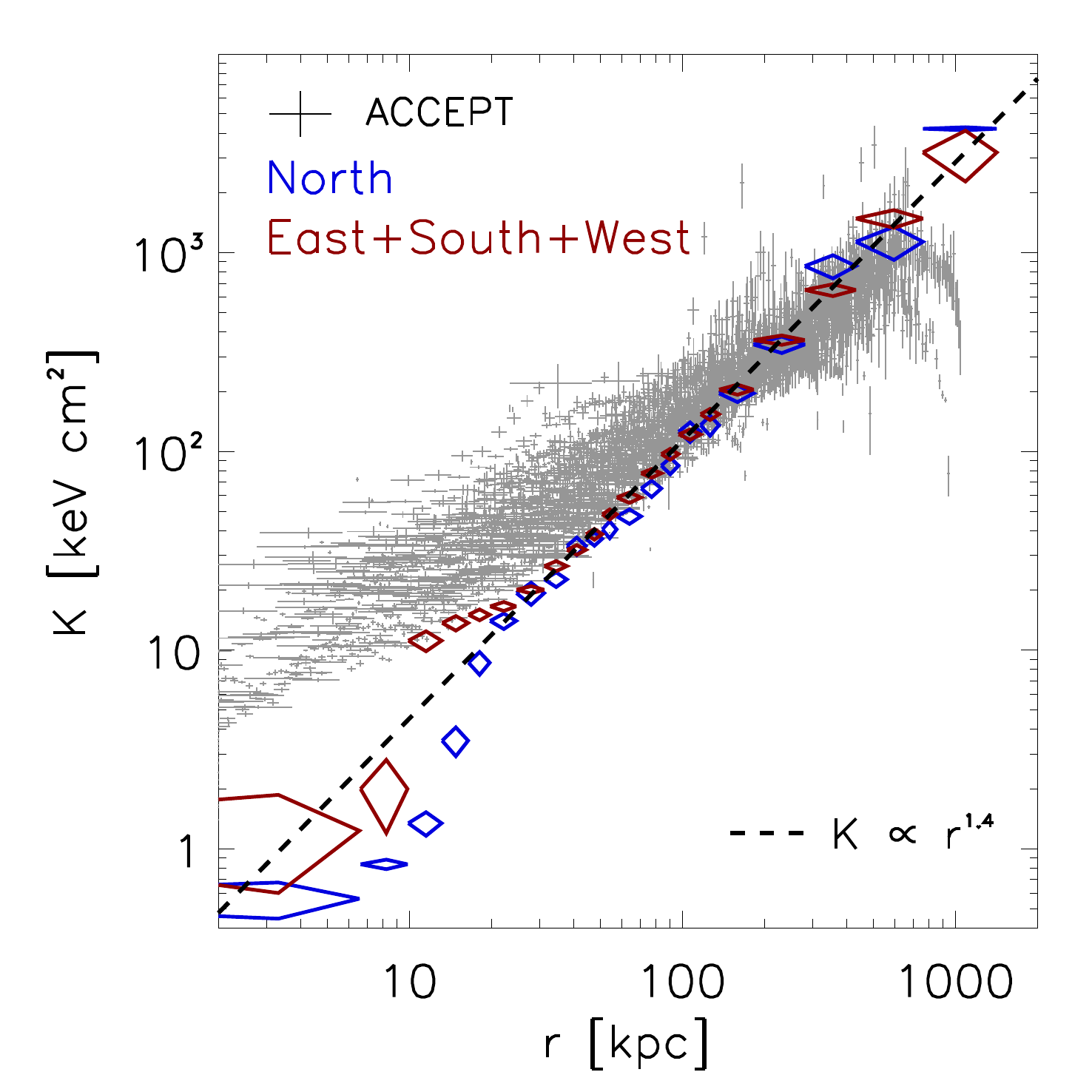}
\caption{Three-dimensional entropy profile extracted along a northern wedge (90$^{\circ}$ opening angle, tilted west by 10$^{\circ}$ to match the position angle of the jet) and along the remaining 270$^{\circ}$. These profiles show that the area surrounding the low-entropy filament in Figure \ref{fig:specmaps} is similar to a ``typical'' cool core cluster, with a shallower entropy profile between $10 < r < 50$\,kpc than at larger radii. At $r<10$\,kpc, and along the northern filament, the entropy profile drops dramatically, indicating the onset of thermal instabilities.
}
\label{fig:wedgeprofs}
\end{figure}

Given that the low entropy gas is highly concentrated in a filament to the north of the central galaxy, our prior assumption of circular symmetry when making thermodynamic profiles is unjustified. To investigate the azimuthal-dependence of the thermodynamic profiles, we make two new sets of profiles, one a 90$^{\circ}$ wedge extending north and aligned with the radio jet and enclosing the low-entropy filament, and the other a 270$^{\circ}$ wedge extending in all other directions and avoiding the low-entropy filament. We measure three-dimensional profiles following \S2.2.2, where we assume circular symmetry in each of the wedges to simplify the model projection. The resulting entropy profiles are shown in Figure \ref{fig:wedgeprofs}. It is clear from these profiles, along with the spectral maps in Figure \ref{fig:specmaps}, that the cool core can be divided into three regions. Over the bulk of the cool core ($r\lesssim50$\,kpc), the entropy profile looks similar to most clusters, with a break in the entropy profile as seen in \cite{babyk18}. The Phoenix cluster traces the lower envelope of clusters from the ACCEPT sample \citep{cavagnolo09}, but does not appear to be a strong outlier in shape. 
In the inner $\sim$20\,kpc, along a northern filament, 
the gas appears to be condensing, resulting in a steep entropy gradient at $r\sim10$\,kpc.  The azimuthal entropy structure at that radius is highly asymmetric, exhibiting large differences in entropy along different directions from the center.  At smaller radii the distribution of multiphase gas is more uniform in azimuth, and ambient gas in all directions has lower temperature, lower entropy, and $t_{cool}/t_{ff}$  approaching unity.
At these small radii our sampling becomes extremely coarse (10\,kpc = 3 ACIS-I pixels), so it is possible that the asymmetric cooling extends all the way to the center of the galaxy.
 
The thermodynamic maps and directional profiles suggest a cool core that has become prone to multiphase condensation that is occurring primarily along a $\sim$20\,kpc filament oriented in the same direction as the radio jets and the cool gas. While there is a chicken-and-egg question with the cool [O\,\textsc{ii}] and the low-entropy filament -- the fact that they are cospatial can can be explained via cooling of the ICM or mixing of already-cool gas (heating) with the ambient ICM -- the situation with the radio emission is more straightforward. It is hard to imagine a scenario where cooling along a given direction would necessarily produce jets along the same direction, given the vastly different scales between this cooling filament and the orientation of the accretion disk. It is much more plausible that the coincidence of the cool filament and the radio jets/bubbles are indicating that cooling is stimulated preferentially along this direction by a jet. We will investigate this possibility further in the discussion section.


\section{Discussion}

\subsection{Cooling Flows, Star Formation, and Mixing}

The early cooling flow models predicted that gas ought to flow steadily inward as it cools, maintaining a power-law entropy profile and a constant cooling rate within each shell \citep[e.g.,][]{nulsen82,fabian84}. We know now that
the simple, steady, pure-cooling model does not apply to the vast majority of cluster cores,
presumably 
due to heat input from black-hole accretion that compensates for radiative cooling (i.e., AGN feedback).
Here, we reconsider these early cooling flow models in the context of modern data, specifically comparing to these new observations for the Phoenix cluster.
Following \cite{voit11}, we combine $\dot{M}(r) = 4\pi r^2 \rho \frac{dr}{dt}$ with $\frac{d\ln K}{dt} = -1/t_{cool}$ to obtain:   

\begin{equation}
\dot{M}(r) = \frac{4\pi r^3 \rho}{\alpha t_{cool}} ~.
\label{eq:mdot}
\end{equation}

\begin{figure}[t!]
\centering
\includegraphics[width=0.49\textwidth]{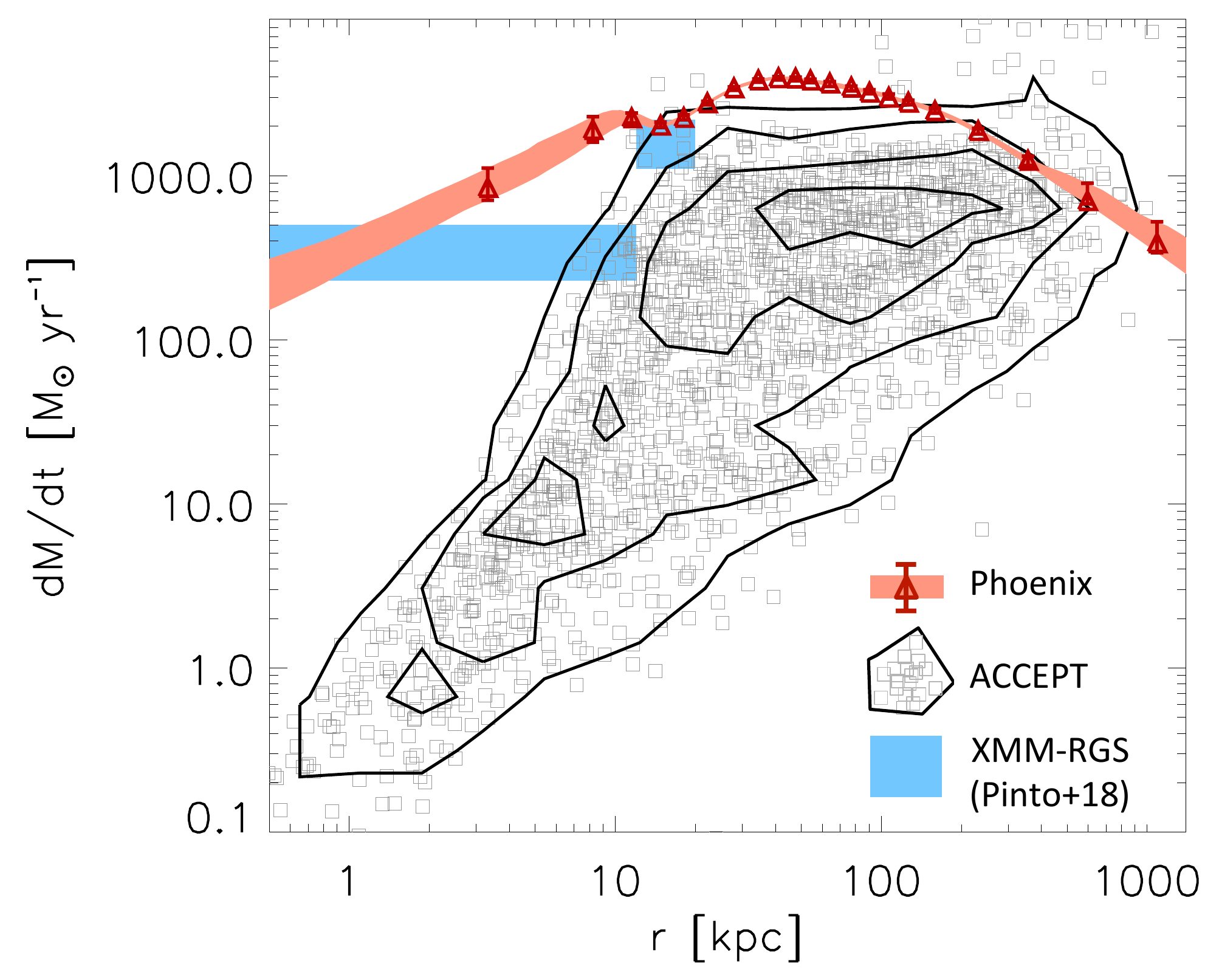}
\caption{Classical cooling rate as a function of radius, defined following Equation \ref{eq:mdot}. Grey squares show profiles for 91 cool core clusters in the ACCEPT sample \citep{cavagnolo09}, resampled to a common binning. Black contours show the distribution of these profiles, which is relatively flat at $r\gtrsim30$\,kpc, and steeply declining at interior radii. This rapid decline in the cooling rate at the center of the cluster is prime evidence that AGN feedback is effectively preventing cooling at these radii. Red points and the shaded red curve show the cooling rate profile for the Phoenix cluster, which is significantly flatter than the median cool core profile, with a central cooling rate elevated by a factor of $\sim$10$^3$ above the typical low-$z$ cluster. Blue shaded regions show the XMM-RGS spectroscopic cooling rates from \cite{pinto18}, where we have converted temperature bins into radial bins using the three-dimensional temperature profile from Figure \ref{fig:xray_profs}. These data further support the picture in which cooling is proceeding with relatively little impedance at the center of the Phoenix cluster.}
\label{fig:dmdt}
\end{figure}

\noindent{}where $\alpha = \frac{d\ln K}{d\ln r}$. 
This is essentially identical to Equation 12 of \cite{voit11}, but in a more convenient form for application to observations. In a homogeneous steady-state cooling flow, the factor $\alpha$ implicitly accounts for gravitational work done on the cooling gas. 
This equation provides the mass inflow rate as a cooling shell passes through radius $r$ with speed $v \equiv -(\frac{d\ln K}{dr} \times t_{cool})^{-1}$. 
Using Equation \ref{eq:mdot}, we calculate the inferred steady cooling rate as a function of radius for 91 cool core clusters in the ACCEPT sample \citep{cavagnolo09}, where we have re-fit the entropy profile with a broken power law following \cite{babyk18}. These profiles are shown in Figure \ref{fig:dmdt}, and exhibit a fairly general behavior. At large radii, the points occupy a locus at  $\dot{M} \sim 600$ M$_{\odot}$ yr$^{-1}$, with no strong radial dependence. This suggests that, at radii $\gtrsim$30\,kpc, there is no strong evidence against gas cooling and flowing inwards. At small radii, however, the classical cooling rate is highly suppressed, by 2--3 orders of magnitude, as has been seen for decades in the centers of cool core clusters \citep[see e.g.,][]{voigt04, fabian12}. This is consistent with X-ray spectroscopic observations, in which significant cooling is observed at high temperature (i.e., 4$\rightarrow$2\,keV) with only upper limits at low temperature \citep[i.e., 1$\rightarrow$0.5\,keV;][]{peterson06}. The implication of Figure \ref{fig:dmdt} is that cooling in the inner $\sim$50\,kpc is being regulated by AGN feedback, leading to highly suppressed cooling rates in the inner $\sim$10\,kpc and, in turn, highly suppressed star formation rates in the central galaxy \citep[see][for a recent summary]{mcdonald18a} 

We also show in Figure \ref{fig:dmdt} the inferred cooling profile for the Phoenix cluster. This profile is relatively flat over 3 decades in radius, with a median value of $\sim$3000 M$_{\odot}$ yr$^{-1}$ (see also Figure \ref{fig:kprof}). At small radii ($r<10$\,kpc), the profile dips down to $\sim$800 M$_{\odot}$ yr$^{-1}$, which is a relatively small change compared to the 2--3 orders of magnitude that most clusters decline at similar radii. We compare this profile to estimates of the cooling rate from XMM RGS spectroscopy \citep{pinto18}. Based on Figure \ref{fig:xray_profs}, we estimate that the cooling from $2\rightarrow0$\,keV is localized to $r<12$\,kpc, and that the gas cooling from $4\rightarrow2$\,keV is in a shell at $12 < r < 20$\,kpc. 
These data are within a factor of a few of the maximum cooling rate, providing additional evidence in support of a weakly-suppressed cooling flow. 

In the Phoenix cluster, at low temperatures ($<$2\,keV), \cite{pinto18} measure an X-ray spectroscopic cooling rate of 350$^{+150}_{-120}$ M$_{\odot}$ yr$^{-1}$. This can be compared to star formation rates measured in a variety of ways. \cite{tozzi15b} model the far-IR SED of the central galaxy and find a star formation rate of $530 \pm 50$ M$_{\odot}$ yr$^{-1}$, while \cite{mcdonald15b} use a combination of far-UV and optical spectroscopy to arrive at a total star formation rate of $610 \pm 50$ M$_{\odot}$ yr$^{-1}$. \cite{mittal17} combine all available data spanning rest frame 1000--10000\AA\ to arrive at a star formation rate of 454--494 M$_{\odot}$ yr$^{-1}$, which is based only on the inner 20\,kpc -- a relatively conservative aperture correction would increase this to $\sim$550 M$_{\odot}$ yr$^{-1}$ \citep{mcdonald15b}. All three of these estimates of the star formation rate, which represent our best estimates to date, are consistent with one another and with the cooling rate derived from X-ray spectroscopy \citep{pinto18}, at the $\sim$1.5$\sigma$ level. Both \cite{mcdonald15b} and \cite{mittal17} rely heavily on well-motivated extinction corrections, which are factors of several, while \cite{tozzi15b} relies on a careful removal of the contribution to the far-IR luminosity from the central dust-obscured QSO. 

The combination of the star formation rates, XMM RGS data \citep{pinto18}, and the classical cooling flow model (Figure \ref{fig:dmdt}), all suggest that cooling is suppressed by a factor of $\sim$5 in the inner $\sim$10\,kpc of the Phoenix cluster, from rates of 2000--3000 M$_{\odot}$ yr$^{-1}$ at larger radii and higher temperature. While significant, this is much less than the typical factor of $\sim$100 found in nearly every other cluster, which led to the formation of the cooling flow problem \citep[see e.g.,][for a recent summary of the literature]{mcdonald18a}. This factor of $\sim$5 reduction of the cooling rate could be provided by the central AGN, which is outputting 1.7--7.2 $\times$ 10$^{45}$ erg s$^{-1}$ in energy \citep{hlavacek15,mcdonald15b} in the inner $\sim$20\,kpc -- exceeding the total cooling luminosity over the same volume.

The biggest inconsistency in this picture is the luminosity of the O\textsc{vi}~$\lambda\lambda$1032,1038 doublet (which probes gas at $\sim$10$^{5.5}$\,K) as measured using the HST Cosmic Origins Spectrograph and reported in \cite{mcdonald15b}. The reported luminosity of $7.55\times10^{43}$ erg s$^{-1}$ implies a cooling rate of 55,000 M$_{\odot}$ yr$^{-1}$, which is considerably more than any of the other cooling or star formation rates. This luminosity is corrected for extinction based on the young stellar populations, which is likely an over-correction if the 10$^{5.5}$\,K gas lies outside of the star forming regions. Assuming no extinction -- which represents an under-correction -- implies a luminosity of L$_{OVI} = 1.1 \times 10^{43}$ erg s$^{-1}$. In a simple picture where the dust is contained within a shell of cooling gas, roughly half of the O\textsc{vi} flux will be extincted, so a reasonable correction would be roughly a factor of 2, or L$_{OVI} = 2 \times 10^{43}$ erg s$^{-1}$. The implied cooling rate, using \textsc{cloudy} models \citep{chatzikos15}, is $\sim$15,000 M$_{\odot}$ yr$^{-1}$ -- still a factor of $\sim$50 higher than the measured cooling rate at low temperatures \citep[for a more detailed discussion, see][]{mcdonald15b}.  \cite{pinto18} showed that the observed \textsc{Ovi} flux is consistent with an additional source of photoionization with $\sim$1 erg s$^{-1}$ cm and a powerlaw spectrum similar to those observed in AGN. If this is indeed the case, one would expect an $r^{-2}$ dependence to the \textsc{Ovi} flux. However, given the limited angular resolution of the HST-COS data, such measurements are not possible.

An alternative origin for the \textsc{Ovi} emission is mixing layers. \cite{shelton18} show that, for a small ($<$1\,kpc), cool ($<$10$^4$\,K) sphere falling through a hot (10$^6$\,K), low-density ($10^{-4}$ cm$^{-3}$) medium, one would expect an \textsc{Ovi} luminosity of $\sim 5 \times 10^{34}$ erg s$^{-1}$. Scaling this by a factor of 10$^6$ to account for the significantly higher density in the ICM compared to these simulations, coupled with the fact that the \textsc{Ovi} emission goes like the density squared yields an expected \textsc{Ovi} luminosity of $5\times10^{40}$ erg s$^{-1}$ for a $5\times10^4$ M$_{\odot}$ cool cloud mixing with a hotter medium. 
For a total luminosity of L$_{OVI} = 2 \times 10^{43}$ erg s$^{-1}$, this would imply 400 mixing clouds, or $2\times10^7$ M$_{\odot}$ of cool material mixing with the hotter medium. The fact that we detect $\sim$1000 times more cool gas than this implies that mixing of the hot and cool media could more than account for the observed \textsc{Ovi} flux. Given the huge extrapolations in going from the simulated geometry, temperatures, and densities to those in the Phoenix cluster, this comparison is meant to be illustrative and not definitive. 

In summary, the deep \emph{Chandra} and \emph{XMM-Newton} data, combined with a variety of star formation estimates all support a picture in which $\sim$500 M$_{\odot}$ yr$^{-1}$ is condensing out of the hot phase and forming stars in the inner $\sim$10\,kpc of the Phoenix cluster. This implies that cooling is only weakly suppressed in the core of the Phoenix cluster, and that it is the closest system to a steady, homogeneous cooling flow that we have yet discovered. The high O\textsc{vi}~$\lambda\lambda$1032,1038 luminosity, coupled with other high-ionization lines \citep{mcdonald14a}, implies that either the central AGN is providing significant ionization on $\sim$10\,kpc scales or,  perhaps more likely, that the cool gas is mixing with the hot ambient medium as it falls into the center of the cluster.

\subsection{Compton Heating \& Cooling}

\begin{figure}[b]
\centering
\includegraphics[width=0.49\textwidth]{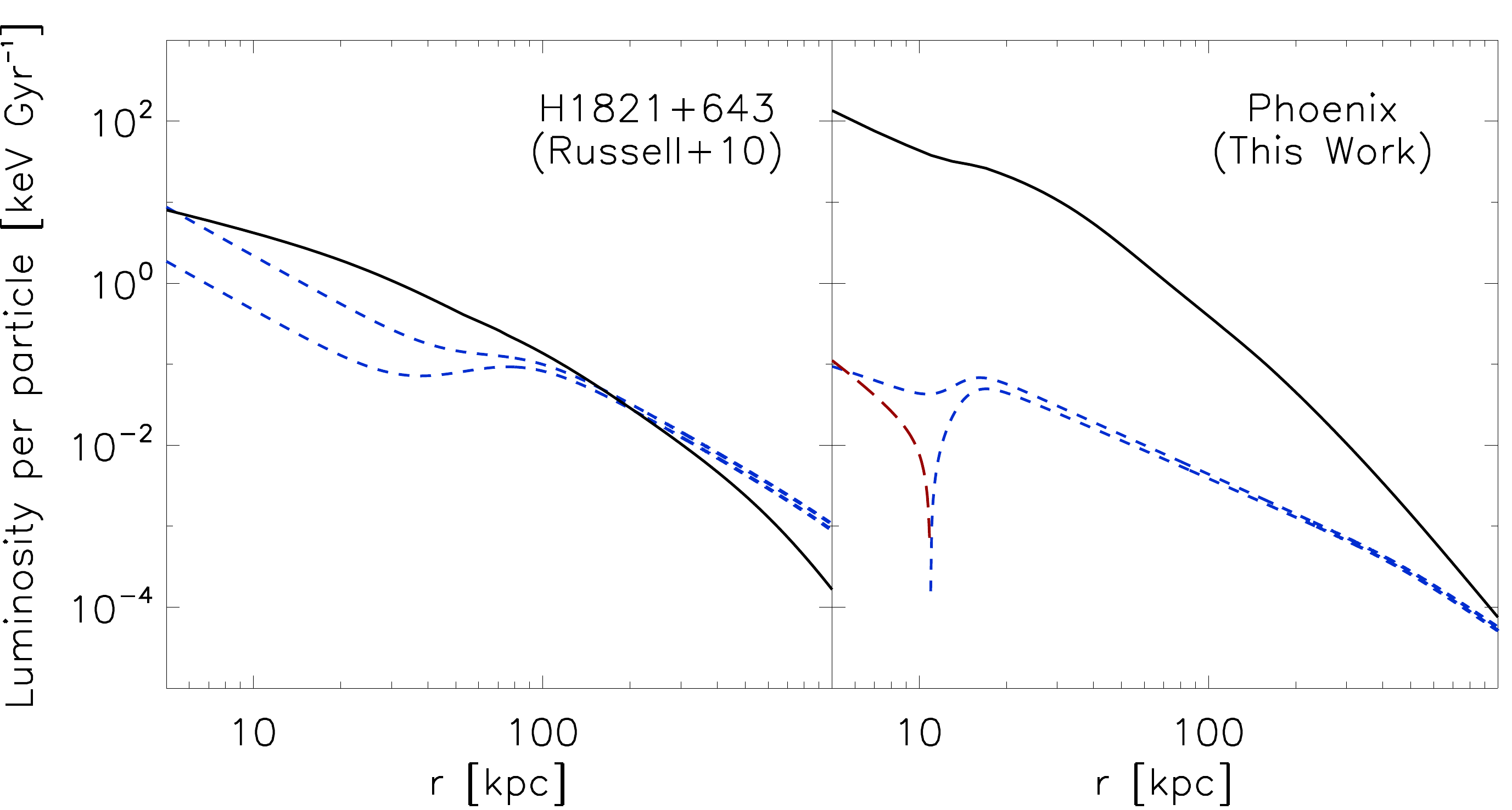}
\caption{Luminosity (cooling rate) per particle as a function of radius for the H1821+643 (left) and Phoenix (right) clusters. In black, we show the cooling rate due to thermal bremsstrahlung emission, based on the measured density and temperature profiles. The dotted blue (red) curves show the Compton cooling (heating) rate as a function of radius, where the upper curve is for a typical unobscured QSO spectrum, while the lower curve is for a typical obscured QSO spectrum \citep{sazanov04}. This figure demonstrates that, while Compton cooling is important for the lower-mass H1821+643 cluster, it is likely not contributing in a meaningful way to the thermodynamics at the center of the Phoenix cluster.
}
\label{fig:compton}
\end{figure}


An uncommon feature of the Phoenix cluster is that it has a bright central X-ray point source, implying a rapidly accreting ($\gtrsim 0.1 \dot{M}_{Edd}$) central supermassive black hole.
There are only three other systems hosting central AGN that are accreting near the Eddington rate, at which point radiative feedback may be relevant to the heating/cooling balance \citep{russell10}.
Given that the characteristic Compton temperature is 0.7\,keV (1.7\,keV) for an unobscured (obscured) quasar, from \cite{sazanov04}, we expect the radiation to effectively cool the surrounding plasma which ought to be hotter than the Compton temperature \citep[e.g.,][]{crawford91}. 
This may not be the case in Phoenix, where the central AGN is heavily obscured (see Figure \ref{fig:nucspec}).

We estimate the relative contributions to the energy gain/loss from Compton cooling/heating and bremsstrahlung cooling as:

\begin{equation}
\dot{E}_{Comp} = \frac{L_{X,bol}\sigma_T}{4\pi r^2} \frac{4k_B(T_C-T)}{m_ec^2}
\end{equation}
\begin{equation}
\dot{E}_{Brem} \sim \Lambda(T,Z)n_e^2
\end{equation}

\noindent{}where $T_C$ is the characteristic temperature of the QSO emission and we assume the cooling function $\Lambda(T,Z)$ from \cite{sutherland93}. We compare the relative contribution of these two processes as a function of radius in Figure \ref{fig:compton} for the Phoenix cluster, and for H1821+643, which has a central QSO that is roughly 30 times more luminous than in Phoenix \citep{russell10}. For H1821+643, the central QSO is unobscured, which implies that Compton cooling probably dominates over bremsstrahlung cooling in the inner few kpc, and is contributing at a similar level over a large range in radii. For the Phoenix cluster, Compton cooling is contributing at a level $\sim$2--3 orders of magnitude less than bremsstrahlung and, in the innermost region, an obscured quasar would lead to heating, rather than cooling. The vast difference between these two systems is due to the fact that the quasar in H1821+643 is a factor of $\sim$30$\times$ more luminous than in Phoenix, while the ICM in Phoenix is roughly an order of magnitude more dense, leading to a much stronger contribution from bremsstrahlung cooling.

In order for Compton cooling to be contributing at the same level as bremsstrahlung in the inner region of the Phoenix cluster, the central point source would need to be a factor of $\sim$1000$\times$ more luminous, which would imply an accretion rate 150$\times$ the Eddington rate. Thus, we can conclude that, while important for H1821+643 and potentially other low-mass systems with highly-luminous central AGN, Compton cooling/heating is unlikely to have played an important role in the thermodynamics in the cool core of the Phoenix cluster.

\subsection{Reconciling Observations of Both Positive and Negative AGN Feedback in Phoenix}

Figures \ref{fig:nHzmap}, \ref{fig:specmaps}, and \ref{fig:wedgeprofs} suggest that the lowest entropy gas in the core of the Phoenix cluster is aligned along the north-south direction, and is the likely origin of the cool ($\lesssim$10$^4$\,K) gas that is fueling star formation. In Figure \ref{fig:multiband}, we show that this is also the direction of the radio jets. Given how well-matched the morphologies of the cool gas, X-ray cavities, and radio jets are, it is unlikely that this spatial correspondence is a coincidence. We also find it unlikely that the jets are aligned with the cooling axis as a reactionary effect -- there is no reason to believe that the accretion disk of the central black hole would be aligned perpendicular to this cooling axis and launch jets preferentially back towards the origin of its fuel supply. This leaves, as the most favorable explanation for the alignment of the radio jets and the cooling axis, the possibility that these jets are stimulating cooling along their path. 
Here, we investigate in detail how the radio-loud AGN in the Phoenix cluster promotes multiphase condensation of intracluster gas while simultaneously providing the heat input necessary to prevent cooling that would be even more catastrophic.

\subsubsection{Negative Feedback (Heating): Bubbles and Shocks}

\begin{figure}[t!]
\centering
\includegraphics[width=0.49\textwidth]{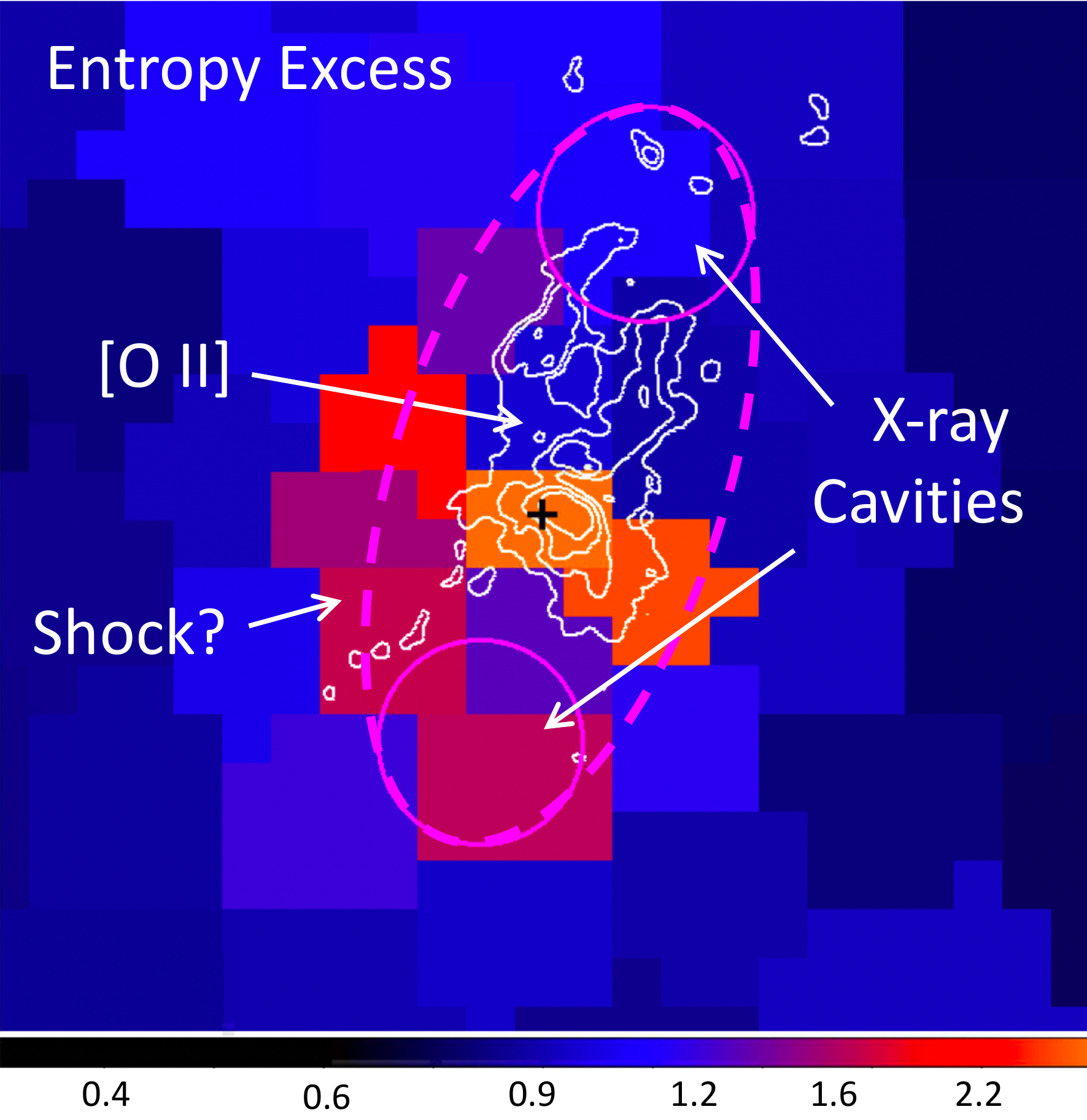}
\caption{Excess entropy map, produced by dividing the two-dimensional entropy map (Figure \ref{fig:specmaps}) by the prediction for a steady, homogeneous cooling flow (see Figure \ref{fig:kprof}). Overplotted are the X-ray cavities (magenta circles), the location of the X-ray AGN (black cross), and the highest surface brightness [O\,\textsc{ii}] emission (white contours). We also show, for illustrative purposes, a magenta ring at the location that we think there may be a cocoon shock. This ring is coincident with a ring of high-entropy gas, where the entropy (and cooling time) is elevated compared to the predictions from the simple cooling flow model.
}
\label{fig:excess_entropy}
\end{figure}

The X-ray residual map (Figure \ref{fig:multiband}) shows two clear cavities, individually detected at a significance of $>$6$\sigma$, which were previously reported by \cite{hlavacek15} and \cite{mcdonald15b}. Both of these cavities are colocated with radio jets, as one would expect if they were bubbles being inflated in the ICM by mechanical feedback. We calculate the mechanical power of these jets following \cite{birzan04} and \cite{rafferty06}. We measure the total pressure, $P_T \sim 2n_e kT$, at the locations of the northern and southern bubbles using the wedge profiles shown in Figure \ref{fig:wedgeprofs}. This is then used to calculate the total enthalpy of the two bubbles shown in Figure \ref{fig:multiband}, assuming $E_{cav}=4P_TV$, where $V$ is the volume assuming a spheroid geometry. We estimate the age of the bubbles assuming the buoyant rise time $t_{buoy} \sim R\sqrt{SC/2gV}$, where $R$ is the distance from the central AGN to the center of the bubble, $S$ is the cross-section of the bubble, $C$ is the drag coefficient \citep[$C=0.75$;][]{churazov01}, $V$ is the volume of the bubble, and $g$ is the local gravitational acceleration. The latter is related to the free fall time by $t_{ff} = \sqrt{2r/g}$. Combining these, and defining $P_{cav} = E_{cav}/t_{buoy}$, yields a total mechanical power of $1.0_{-0.4}^{+1.5} \times 10^{46}$ erg s$^{-1}$, with the uncertainty dominated by the uncertainty in the cavity geometry. This power is sufficient to offset cooling in the inner $\sim$100\,kpc, within which we measure an unobscured cooling luminosity of $L_{cool} = 1.1 \pm 0.1 \times 10^{46}$ erg s$^{-1}$. If we, instead, assume that the bubbles are rising at the sound speed, the total mechanical power is reduced to $0.5_{-0.2}^{+0.6}  \times 10^{46}$ erg s$^{-1}$, which is still consistent with the cooling luminosity, but on the lower side.
Despite this apparent energy balance, we observe an abundance of condensation, as indicated by multiphase gas and the formation of new stars, suggesting that this mechanical energy has yet to couple to the hot ICM, or that (unsurprisingly) the picture is more complicated than simple energy balance.

To further investigate the heating effects of AGN feedback, we generate an ``excess entropy'' map, by dividing the two-dimensional entropy map (e.g., Figure \ref{fig:specmaps}) by the prediction for a steady, homogeneous cooling flow ($K \propto r^{1.4}$; Figure \ref{fig:kprof}). The resultant ratio is shown in Figure \ref{fig:excess_entropy}. This figure shows that the entropy follows the $r^{1.4}$ expectation for a steady-state cooling flow over most of the cool core volume, as depicted by blue in this map. However, there is a partial ring, resolved into 6 spatial regions, surrounding the cluster center with a projected entropy boosted by a factor of $\sim$2. Geometrically, this partial ring is consistent with a cocoon shock enclosing the two bubbles, centered on the AGN (dashed magenta ellipse in Figure \ref{fig:excess_entropy}).  This excess entropy region is coincident with a rise in density ($\sim$1.4$\times$) and projected temperature ($\sim$1.3$\times$), as expected for an outward-moving shock. If this entropy jump is indeed due to a shock, it implies a relatively low Mach number of $\sim$1.3--1.5. However we emphasize that these data are only suggestive, and a significantly deeper exposure would be needed to determine if this is, indeed, a shock.

In summary, there is compelling evidence that feedback from the central AGN is heating the surrounding medium and may be capable of preventing further runaway cooling. The total mechanical power, derived from the extent and location of the X-ray bubbles, is $1.0_{-0.4}^{+1.5} \times 10^{46}$ erg s$^{-1}$, which is sufficient to offset the cooling luminosity of $L_{cool} = 1.1 \pm 0.1 \times 10^{46}$ erg s$^{-1}$. There is some evidence for asymmetric heating in the excess entropy map which may explain why the evidence for cooling that we observe, in the form of low-entropy and multiphase gas, is also highly asymmetric.

\subsubsection{Positive Feedback (Cooling): Uplift and Turbulence}

\begin{figure}[t!]
\centering
\includegraphics[width=0.47\textwidth]{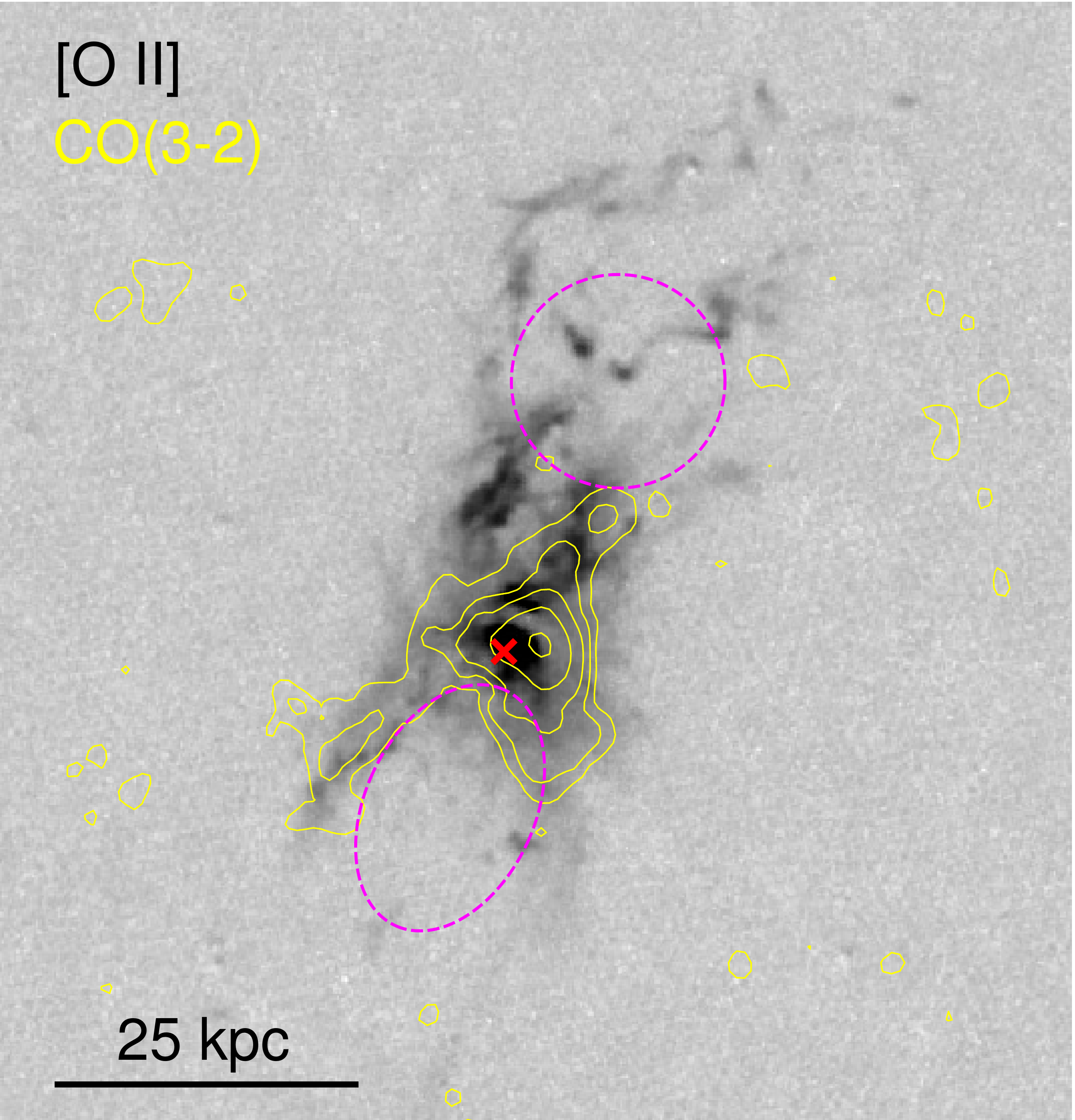}
\caption{[O\,\textsc{ii}] map of the inner $\sim$50\,kpc of the Phoenix cluster. Overlaid on this map are contours showing the CO(3-2) emission, from \cite{russell17}, and the locations of the X-ray cavities, shown as magenta ellipses. This figure demonstrates that the bulk of the cool, multiphase gas is either behind, or draped around, the bubbles. However, a small fraction of the multiphase gas leads the bubbles, extending for $\sim$20\,kpc beyond the leading edge of both the northern and southern bubbles.}
\label{fig:cavims}
\end{figure}

In Figure \ref{fig:cavims} we summarize the correspondence between the central AGN, the X-ray cavities, which define the jet direction, the cool [O\,\textsc{ii}]-emitting gas, and the molecular CO(3-2) emission in the inner $\sim$50\,kpc of the Phoenix cluster.  As has been mentioned here and in previous works \citep[e.g.,][]{russell17}, there is a strong morphological connection between the multiphase gas and the X-ray bubbles. Both the southern and northern bubbles appear to have multiphase gas draped around the trailing edges, with the bulk of the cool gas lying behind the northern bubble. This seems to imply, counterintuitively, that the strong mechanical feedback appears to be having a net positive (cooling) effect on the gas in its path. We investigate possible explanations for this below.

Considering the entropy profiles in the jet and anti-jet direction (Figure \ref{fig:wedgeprofs}), the two profiles diverge at $\sim$30\,kpc. This radius corresponds to the leading edge of the northern bubble. Interior to this radius, cooling appears to be more effectively suppressed in the anti-jet direction, while along the jet direction cooling is intensified (Figure \ref{fig:excess_entropy}). 
This is consistent with a picture in which jet-inflated bubbles promote multiphase condensation in their wake, as low entropy gas is lifted behind buoyantly-rising bubbles, leading to condensations when buoyancy is no longer able to suppress the development of density contrasts \citep[e.g.,][]{brighenti15}. 
The morphology of the [O\,\textsc{ii}]-emitting gas to the north of the central AGN supports this picture (Figure \ref{fig:cavims}). This cool gas is reminiscent of the ``horseshoe'' filament in NGC1275 \citep{conselice01}, which in turn resembles simulations of gas condensing behind a rising bubble \citep{fabian03,reynolds05, brighenti15}. In this scenario, we would expect to see a cool wake behind the bubble, as low-entropy gas is dredged up from the center to fill the void left behind the buoyant bubble. Indeed, this low-entropy ``bridge'' between the nucleus and the bubble is observed (Figure \ref{fig:specmaps}). Based on Archimedes' principle, the maximum amount of low entropy gas that can be uplifted behind the bubble is equal to the amount of gas displaced from the bubble. Assuming the bubble geometries shown in Figure \ref{fig:cavims}, this corresponds to $1.9_{-0.6}^{+3.8} \times 10^{10}$ M$_{\odot}$ and  $3.8_{-1.3}^{+3.7} \times 10^{10}$ M$_{\odot}$ for the northern and southern bubbles, respectively. This is significantly more than the total mass in cold molecular gas \citep{russell17}, and also exceeds the inferred mass of the X-ray-absorbing cool gas (\S3.1). Likewise, if we integrate the mass of the low entropy filament to the north, assuming a cylindrical geometry that is 20\,kpc long and 10\,kpc in diameter, we infer a mass of $\sim0.5\times10^{10}$ M$_{\odot}$. Thus, it is completely plausible that all of the cool, multiphase gas in the inner $\sim$20\,kpc is a result of uplift caused by the buoyantly-rising bubbles effectively increasing the freefall time of the low-entropy gas, making multiphase condensation inevitable when $t_{cool}/t_{ff} \sim 1$ \citep{revaz08,li14,mcnamara16}. While this mechanism can explain the vast majority of the cool gas, and almost all of the observed star formation and molecular gas, it cannot explain the full extent of the emission line nebula, which reaches well beyond the leading edge of both sets of cavities (Figure \ref{fig:cavims}).

In addition to ordered uplift due to the buoyant bubbles, we expect significant turbulence due to the combination of AGN feedback and the accretion history of the cluster. The presence of turbulence 
should enhance cooling \citep[e.g.,][]{gaspari18,voit18b}. \cite{gaspari14b} show, via hydrodynamical simulations, that the ICM velocity and density power spectra are closely related, meaning that an increase in the overall level of turbulence should correspond to an increase in the density contrast $\delta\rho/\rho$. Given that the cooling rate is proportional to the square of the local density, an overall increase in the level of turbulence should lead to more rapid condensation of the hot gas. \cite{voit18b} provided a simplified, analytic description of this process, demonstrating that if the level of turbulence in the ICM reaches $\sim$50\% of the stellar velocity dispersion of the central galaxy, the turbulent motions can be strong enough to counteract the tendency for buoyancy to suppress condensation. 

Following \cite{gaspari18}, we assume that the bulk motions of the cool gas trace the level of turbulence in the hot gas, akin to a flag blowing in the wind. Based on the kinematics of the [O\,\textsc{ii}] emission line \citep{mcdonald14a}, we measure a line-of-sight velocity dispersion of $\sim$250\,km/s, or a 3-dimensional velocity dispersion of $\sigma_{3D} = \sqrt{3}\sigma_{1D} \sim 430$\,km/s. This inferred level of turbulence is a factor of $\sim$2 higher than that directly measured in the Perseus cluster \citep{hitomi16}, and roughly equal to the expected stellar velocity dispersion of the central galaxy. Such high turbulent velocities, if real, may provide an important clue to why cooling is so effective in the Phoenix cluster. In the future, with calorimeters on the XRISM, Athena, and Lynx missions, we will be able to look for correlations between the level of turbulence and condensation. The Phoenix cluster would provide an excellent test-bed for theories which posit that turbulence plays a critical role in the feedback/cooling cycle \citep[e.g.,][]{gaspari17,voit18b}.

In summary, while there is sufficient mechanical energy in the radio jets to offset all of the radiative losses, the effects of this outburst appear to be strongly asymmetric. Along the jet direction, specifically the northern jet, condensation appears to be enhanced. This is likely due to a combination of i) the buoyant uplift of large amounts of low entropy gas behind rising bubbles to radii where $t_{cool}/t_{ff} \sim 1$, and ii) an overall increase in the turbulent velocities of the ICM, which ought to promote condensation. These processes may have seeded an initial condensation event, which may then become amplified as the local turbulence is further enhanced due to the massive starburst.  Radiative losses from this highly multiphase gas may also be boosted if mixing of the hot and cold gas can produce mixing layers at 10$^5$--10$^6$\,K, near the peak of the cooling curve. In the opposite direction, the entropy of the gas is significantly increased,  perhaps due to a cocoon shock surrounding the bubbles. The net effect of this asymmetric feedback is that, as a whole, cooling is only weakly suppressed.

\subsection{Saturating AGN Feedback}

The data presented here, and in earlier works, have demonstrated that the Phoenix cluster is cooling near the predicted rate for a homogeneous, steady cooling flow. This cooling is homogenous on large scales, and becomes asymmetric in the inner $\sim$20\,kpc, where the influence of the central AGN appears to be aiding cooling in one direction and hindering it in another direction. 
Yet, the question remains: why is this system, of all of the known clusters, cooling at such a prodigious rate?

In \cite{mcdonald18a}, we offered a potential explanation, which is connected to the high mass of the Phoenix cluster. For the most massive galaxy clusters, we expect the central supermassive black hole to be undersized compared to the mass of the cool core. This can be understood by considering the outcome of a cluster-cluster or cluster-group merger. As a smaller halo falls into the main halo, 
its lowest-entropy gas will naturally sink towards the center. Thus, in the absence of strong shocks or a significant boost in turbulence, both which could raise the entropy during a merger, 
the total mass of low-entropy gas in the cool core will increase when two clusters merge. On the other hand, there is no requirement that the central galaxies merge on equally-short ($\sim$1\,Gyr) timescales, as dynamical friction can require $\sim$10 Gyr to slow down the most massive galaxies and cause them to spiral in towards the center \citep{sarazin86}. Thus, while the amount of low-entropy gas will inevitably increase during an accretion event, the mass of the central galaxy and, thus, its supermassive black hole is not required to grow at the same rate. For clusters that grow rapidly to very high masses, this ought to lead to undersized supermassive black holes at the centers of clusters.

\cite{russell13} show that, for supermassive black holes accreting near the Eddington rate, the mechanical power plateaus at $\sim$10$^{-2}$ L$_{Edd}$ while the radiative power continues to rise with increasing accretion rate. For a 10$^9$ M$_{\odot}$ black hole, this corresponds to a mechanical power of $\sim$10$^{45}$ erg/s, while for a 10$^{10}$ M$_{\odot}$ black hole, which is the expectation for the central black hole in a rich cluster, the limiting mechanical power would be $\sim$10$^{46}$ erg/s. If the Phoenix cluster grew its cool core rapidly, so that its central galaxy and accompanying black hole were initially undersized compared to the cluster, it would be unable to provide the necessary 10$^{46}$ erg/s in mechanical power to regulate cooling. At the current accretion rate \citep[$\sim$60 M$_{\odot}$ yr$^{-1}$;][]{mcdonald12c}, the black hole would grow to 10$^{10}$ M$_{\odot}$ in $\sim$200 Myr, which would be enough time for the cooling flow to develop but not for 
the total accumulated mass in young stars to become overly large. This rapid growth would facilitate a rise in the maximum mechanical power from $\sim$10$^{45}$ erg/s to $\sim$10$^{46}$ erg/s over $\sim$200\,Myr, allowing it to begin to regulate cooling.

This picture is supported by the fact that the most massive relaxed clusters \citep[e.g., Phoenix, Abell~1835, RX~J1347.5-1145][]{mantz16} are also among the most rapidly cooling \citep{mcdonald18a}. To go beyond speculation, we require central galaxy masses or, even better, black hole masses for a large number of clusters spanning a wide range in cooling efficiencies. If mechanical feedback does, indeed, saturate after a period of rapid cluster growth, then we would expect central cluster galaxies and their supermassive black holes to be undersized in clusters for which the central galaxy is rapidly forming stars.

\section{Summary}

In this work, we have presented new data from the \emph{Chandra X-ray Observatory}, \emph{Hubble Space Telescope}, and the Karl Jansky Very Large Array. These data provide an order of magnitude improvement in depth and/or angular resolution at X-ray, optical, and radio wavelengths, yielding a detailed view of the core of the Phoenix cluster. In particular, the deeper X-ray data has allowed a careful modeling of the central point source, allowing thermodynamic properties of the intracluster medium to be probed on scales similar to the central starburst for the first time. Our findings are summarized as follows:

\begin{itemize}

\item We confirm, for the first time, that the previously-reported bubbles in the X-ray \citep{hlavacek15,mcdonald15b} are coincident with radio jets, as mapped by the VLA. These jets appear to be inflating bubbles in the hot ICM, with a total mechanical power of $0.5_{-0.2}^{+0.6} \times 10^{46}$ erg s$^{-1}$ if the bubbles are rising at the sound speed or $1.0_{-0.4}^{+1.5} \times 10^{46}$ erg s$^{-1}$ if they are rising buoyantly. This outburst energy is consistent with balancing the unobscured cooling luminosity in the inner $\sim$100\,kpc of $1.1 \pm 0.1 \times 10^{46}$ erg s$^{-1}$.

\item Narrow-band imaging of the [O\,\textsc{ii}]$\lambda\lambda$3726,3729 doublet with HST has revealed a complex nebula of cool (10$^4$\,K) gas in the inner $\sim$40\,kpc. The morphology of this gas is similar in complexity to that seen in NGC1275 at the core of the Perseus cluster \citep{conselice01}. The bulk of the cool gas traces the locations of previously-detected X-ray bubbles and the newly-detected radio jets, consistent with the picture that this gas is condensing behind buoyant bubbles and falling back onto the central galaxy. Faint filaments of cool gas extend well beyond the leading edge of the bubbles, implying that uplift is not solely responsible for the presence of cool gas.

\item We detect a significant amount of absorption in the X-ray, consistent with total mass of $1.2\times10^{10}$ M$_{\odot}$ of cool material. This gas is spatially extended, consistent in morphology and distribution with the brightest [O\,\textsc{ii}]$\lambda\lambda$3726,3729 and CO(3-2) emission. The cool X-ray-absorbing gas appears to be primarily distributed along a northern filament, lying between the central AGN and the southern edge of the northern bubble, again consistent with an uplift picture. At the measured spectroscopic cooling rate of $\sim$350 M$_{\odot}$ yr$^{-1}$, this much cool gas could accumulate in only $\sim$34\,Myr.

\item We measure a steeply-declining temperature profile, which peaks at $\sim$14\,keV and reaches a projected temperatures as low as 2\,keV at the center. Previous works found a higher central temperature due to contamination from the central X-ray-bright AGN. The three-dimensional temperature profile has an inner temperature of $\sim$1\,keV over a spherical volume 10--20\,kpc in radius. The inferred entropy profile ($K\equiv kTn_e^{-2/3}$) shows no evidence for an entropy excess in the inner region, consistent with a power law with slope $r^{1.4}$ at all radii. Both the azimuthally-averaged temperature and entropy profiles are consistent with predictions from hydrodynamic simulations of pure cooling, or analytic expectations for a homogenous steady-state cooling flow.

\item In the inner $\sim$30\,kpc, the cooling time profile is lower than that measured in any other cluster, reaching a minimum of 10\,Myr at $r\lesssim 5$\,kpc. This is an order of magnitude lower than any other known cluster at similar radius. The ratio of the cooling time to the free fall time, $t_{cool}/t_{ff}$, reaches $\sim$1 in the inner $\sim$5\,kpc and is below 10 in the inner $\sim$30\,kpc. Such low values of $t_{cool}/t_{ff}$ 
remove the buoyancy barrier that limits multiphase condensation in the hot ICM, leading to condensation of the hot gas into cool, star-forming clumps.

\item When we consider the two-dimensional thermodynamic maps, we find that the lowest temperature and lowest entropy material is extended along a $\sim$20\,kpc filament to the north, consistent in location with the cool [O\,\textsc{ii}]-emitting and X-ray absorbing gas. This low-entropy filament, which has $t_{cool}/t_{ff} \sim 1$ along its length, trails behind the northern bubble, consistent with having been uplifted in the rising bubble's wake.  In the off-jet direction, the entropy appears to be increased by a factor of $\sim$2, possibly due to a cocoon shock surrounding the bubbles.

\item We rule out Compton cooling as a contributor in the core of the Phoenix cluster, with thermal bremsstrahlung dominating the cooling at all radii by several orders of magnitude.

\end{itemize}

%
%
%
%

In general, the one-dimensional thermodynamic profiles support a picture in which the Phoenix cluster is cooling near the prediction for a homogenous, steady-state cooling flow. On closer inspection, via two-dimensional maps, this cooling is highly asymmetric, being strongly enhanced along the jet direction and suppressed in the counter direction. The data from \emph{Chandra}, \emph{Hubble}, and the VLA combine to produce a compelling picture in which the central radio-loud AGN is stimulating multiphase condensation in the ICM by uplifting cool, low-entropy gas to radii where it can cool faster than it can fall back, and by generating significant turbulence which will amplify density contrasts and lead to more rapid condensation. The gas that is cooling is likely mixing with the hotter ICM, leading to more rapid cooling and strong high-ionization emission lines, as observed. We propose that the Phoenix cluster may have harbored an undersized central supermassive black hole, which has been unable to provide the necessary feedback to halt cooling. Rapid growth over the past $\sim$200\,Myr may have led to the observed outburst, which appears to be increasing the core entropy. Future observations of the multiphase gas with \emph{Athena}, \emph{Lynx}, and \emph{JWST} will allow the mapping of gas at 10$^5$\,K and 10$^6$\,K, allowing a more complete picture of the multiphase cooling flow in this extreme system.


\section*{Acknowledgements} 
We thank Robin Shelton and Paul Nulsen for illuminating discussions. 
Support for this work was provided to MM and TS by NASA through Chandra Award Number GO7-18124 issued by the Chandra X-ray Observatory Center, which is operated by the Smithsonian Astrophysical Observatory for and on behalf of the National Aeronautics Space Administration under contract NAS8-03060.
Additional support was provided to MM by NASA through a grant from the Space Telescope Science Institute (HST-GO-15315), which is operated by the Association of Universities for Research in Astronomy, Incorporated, under NASA contract NAS5-26555. 
MM also recognizes generous lumbar support from the Adam J.\ Burgasser Endowed Chair in Astrophysics.
MG is supported by the Lyman Spitzer Jr. Fellowship (Princeton University) and by NASA Chandra grants GO7-18121X and GO8-19104X.
RJvW acknowledges support from the VIDI research programme with project number 639.042.729, which is financed by the Netherlands Organization for Scientific Research (NWO). 
The National Radio Astronomy Observatory is a facility of the National Science Foundation operated under cooperative agreement by Associated Universities, Inc.


\end{document}